\documentclass{aa}
\usepackage{times}
\usepackage{graphicx}
\begin{document}

   \thesaurus{06     
              (02.01.2           
               08.09.2 EX\,Dra;  
               08.14.2)}         

\title{Reconstruction of emission sites in the dwarf nova 
       EX\,Draconis\thanks{Based on observations obtained at
                                  the German-Spanish Astronomical
                                  Center, Calar Alto, Spain and at
       Wendelstein Observatory, Germany.}
       }


   \author{Viki Joergens \and Karl--Heinz Mantel \and Heinz Barwig
   \and Otto B\"arnbantner \and Hauke Fiedler}

   \offprints{Viki Joergens (viki@mpe.mpg.de)}

   \institute{{Universit\"ats-Sternwarte M\"unchen,
              Scheinerstr.1, D--81679 M\"unchen, Germany}
%
             }

%
%

   \date{Accepted November 17, 1999}

   \maketitle

   \begin{abstract}
      
      We performed time--resolved spectroscopic \mbox{studies} of the
      double--eclipsing dwarf nova EX\,Dra (formerly HS~1804\,+\,6753) 
      in order to locate line
      emitting sites in the system. Optical spectra recorded during the
      quiescent as well as during the outburst state have been analysed by
      means of Doppler tomography. The computed Doppler images map the 
      system in a variety of emission lines and allow us to compare between  
      \mbox{different} temperatures
      and accretion states. 

      Our studies revealed that the Balmer and 
      \ion{He}{i} emission of
      EX\,Dra during quiescence is mainly formed within a fully established 
      disk and within the gas stream. The Doppler map of H$_{\alpha}$ shows 
      a second emission spot in the accretion disk located far from the  
      region of interaction between the 
     gas stream and the accretion disk.

      We have found a weak hint that secondary star emission contributes
      to the H$_{\alpha}$ line in
      quiescence, obviously caused by photospheric heating due to 
      irradiation by the primary component.
      During
      outburst secondary star emission turns into a very strong
      emission 
      source in the Balmer lines 
      due to the increased accretion rate and an enhanced irradiation by 
      the white dwarf or the boundary layer.
      The Doppler maps of the Balmer and \ion{He}{i} lines during outburst 
      further show emission from the accre\-tion disk. During outburst the 
      gas stream is rarely 
      seen in the Balmer lines but clearly visible in \ion{He}{i} and shows 
      that the disk radius during this high accretion state is about 
      0.2~R$_{L1}$ larger than 
      during the recorded quiescent state.

      The origin of the \ion{C}{ii}
      ($\lambda$\,4267\,{\AA}) line, which is only detectable during eruption 
      can be located by Doppler imaging
      close to the primary component and may therefore be formed in the 
      chromosphere of the white dwarf.

   \end{abstract}

%

\section{Introduction}
\label{Introduction}

The cataclysmic variable \object{EX\,Dra} 
(formerly \object{HS~1804\,+\,6753}), was first
detected in the Hamburger Quasar Survey in 1989 (Reimers D., 1991, 
private comm.).
In the course of follow-up observations the system
turned out to be a double-eclipsing dwarf nova of the U\,Gem class
with a quiescence magnitude of about 14$^m$, a
relatively small outburst amplitude of 1$^m$--2.3$^m$ and an orbital
period of 5.04\,h (\cite{Barwig93}).

Up to now two different approaches have been made to analyse
this binary system. 
Billington~et~al. (1996)
undertook spectroscopic observations
of the red part of the optical wavelength range of EX\,Dra in 1994. 
Analysis of the H$_{\alpha}$ emission by means 
of Doppler tomography 
indicated that the H$_{\alpha}$
emission of EX\,Dra is dominated by the bright spot and the inner
Langrangian region, whereas the accretion disk itself is rarely seen
in H$_{\alpha}$.
Fiedler~et~al. (1997)
used optical spectra covering the spectral
wavelength range between 3500 and 8900\,{\AA} and long term photometric
observations, both recorded during the quiescent as well as the
outburst state, to  
determine the fundamental system parameters. 

Cataclysmic variables (CVs) are close
interacting binary systems containing an accreting white dwarf star
and a Roche lobe filling secondary, typically a main sequence star,
which loses mass via the inner Langrangian point into the Roche lobe of
the white dwarf. Conservation of angular momentum causes the stream
material in non-magnetic CVs
to form a disk around the massive star, which is the
dominating light source of the system
as observations reveal. 
Dwarf novae are a subclass of CVs and frequently show outbursts, which
are episodes of enhanced accretion through the disk and onto the
central object.

Whereas this basic CV
model is well established, the processes leading to mass transfer
through the disk and the transport of angular momentum within the disk
are still outstanding questions.

The energy flux of cataclysmic variables in the optical wavelength
range is dominated by emission from the accretion disk represented by a
continuum flux increasing towards a maximum in the UV and emission lines mostly
originating from
H and He. The line profile shapes depend on the distribution of line
flux over the disk 
and therefore can be used to constrain it. 
Other components of the system, like the two stars, the gas stream and
disk outflows
can also significantly contribute to the emission line flux.
With Doppler tomography, an image reconstruction technique
(\cite{Marsh88}), two--dimensional
images 
of the system in velocity space can be
obtained, which
allow one to ascertain the contributions from the two
stars, the disk and other clear emission sites to the observed line
flux. Doppler mapping can be used as a constraint for \mbox{theories} of line
formation and can indicate the
structure of the disk. The emission distribution over the disk 
in most of the known cataclysmic
variables is far from being uniform or even symmetric.
The interaction between the highly supersonic overflowing secondary
star material and the edge of the rotating disk and the mass transfer
through the disk lead to an inhomogeneous distribution of emission. 
A prominent region of enhanced emission, the  
so--called bright spot is often visible
at the rim of the disk where the stream material impacts 
onto the disk. Recently a group found observational evidence
for spiral structures in an outburst accretion disk by means of Doppler
tomography: 
Steeghs et al. (1997)
found a two armed structure in the 
disk of IP\,Peg during outburst.
Emission from the secondary star can also play a role, as detected for
example for the H$_{\alpha}$ and H$_{\beta}$ line of \object{IP\,Peg} during
quiescence (\cite{Wolf98}).
The knowledge of the distribution of line emission is in addition
crucial for the measurement of the radial velocity of the white dwarf,
since phase-dependent
asymmetries in the emission lines distort
measurements of this parameter.

This paper presents phase--resolved studies of \object{EX\,Dra} by 
analysing the spectra presented in 
Fiedler~et~al. (1997)
in more detail.
Our intention is to locate line emitting sites in
the system and 
to obtain information of the 
line flux distribution over the disk
by means of Doppler tomography.
Spectra recorded
during both the outburst and the quiescent state showing a series of emission
lines allow us to compare Doppler maps of
different temperatures and accretion states to reveal details of
the accretion mechanism in the system EX\,Dra. 

Sect.~\ref{Spectroscopy} describes the time--resolved
spectroscopic 
observations of the eclipsing dwarf nova EX\,Dra taken during the quiescent
as well as during the outburst state and the applied reduction
algorithm. The flux 
calibration of the spectra is presented in Sect.~\ref{Calibration}.
In Sect.~\ref{analyses} the Doppler velocity profiles are discussed and
used to map the emission line regions.
It is followed by a discussion of the images, a comparison between the high
and the low accretion state and a comparison of 
the Doppler map of H$_{\alpha}$ with the
H$_{\alpha}$ map of EX\,Dra performed by 
Billington~et~al. (1996). 
In Sect.~\ref{Conclusions}
the results of this paper are summarized.


\section{Data acquisition and reduction}
\label{Data}

\subsection{Spectroscopy}
\label{Spectroscopy}

The acquisition and reduction of the spectroscopic data 
has already been discussed by Fiedler~et~al. (1997). Therefore in the following only a short summary of the 
spectroscopic data is presented.
A set of 137 
optical spectra of EX\,Dra was recorded in an observing
run at the Calar Alto 3.5\,m telescope with the Cassegrain
Twin Spectrograph in 1992. 
The choosen gratings 
provided a dispersion of 1.7\,{\AA}
per pixel in the blue and 1.1\,{\AA} per pixel in the red spectral
range with a wavelength coverage 
between 3440 and 5330\,{\AA} and 5690 and 6810\,{\AA}
respectively. During these 
observations EX\,Dra was in quiescent state.

An additional sequence of 32 spectra was taken in a second observing 
run at the Calar Alto 3.5\,m telescope in 1993 during which EX\,Dra was found
in (probably an early) outburst state. 
The
wavelength coverage in the red part of the spectra was extended to
longer wavelengths (6200$\dots$8900{\AA}) 
at the cost of the spectral resolution leading to
a dispersion of
2.7\,{\AA} per pixel in the red part of the spectra.
The blue part of the wavelength range (3840$\dots$5630{\AA}) was observed with a dispersion of 1.8\,{\AA} per pixel.

The exposure times varied between 200\,s and 1200\,s.
Spectra of the spectrophotometric standard stars
\object{BD\,+28$^\circ$4211} (Stone standard)
and \object{Wolf\,1346} (Oke standard)
were taken. 

The spectroscopic data were reduced using the so--called Optimal 
Spectrum Extraction Algorithm (\cite{Horne86}), taking into account
bias--subtraction, flatfield--correction, sky--subtraction and
cosmic--ray elimination
and performing a wavelength--calibration.

In order to correct for the wavelength--dependent sensitivity of the
atmosphere 
and of the instruments a separate flux calibration was performed 
(see Sect. \ref{Calibration}).


\subsection[]{Photometry}
\label{Photometry}

Long term photometric observations of EX\,Dra in the quiescent as well as in
the outburst state were performed with
the Multichannel--Multicolour Photometer MCCP (\cite{Barwig87}) attached
to the 80\,cm Wendelstein telescope during the years 1991 to 
1996 and to the 2.2\,m 
telescope at Calar Alto observatory in 1992 and 1993. The data were
recorded with a time resolution of 2\,s and 1\,s. 
The MCCP is a high--speed photometer providing three fiber
channels to measure
the object, a nearby comparison star and the sky background simultaneously.
Atmospheric effects are eliminated using the so--called Standard Reduction
Algorithm
(\cite{Barwig87}) which subtracts the sky background of each of the
five UBVRI colour channels from object and comparison star and divides
the object by the comparison star measurements afterwards. The simultaneous
observing technique in combination with
the reduction algorithm allows to perform photometric measurements even 
under non--photometric conditions. A detailed journal of the photometric
observations from 1991 to 1996 is given by Fiedler et al. (1997).
The B and R light curves from these photometric measurements were used
for the flux 
calibration of the spectra (see Sect.~\ref{Calibration}).


\subsection{Flux calibration of the spectra}
\label{Calibration}

A flux calibration is required in order to account for the wavelength--
and time--dependent sensitivity of the atmosphere, for the
wavelength--dependent sensitivity of the telescope, the spectrograph
and the detector 
and to transform the recorded flux
distribution to an absolute scale.

Under stable atmospheric conditions the flux calibration can be
performed by using recorded spectra of spectrophotometric standard stars,
or alternatively by using simultaneously recorded broadband photometry.
The latter can also be applied under variable atmospheric conditions, when
an instrument like the MCCP is used which allows to obtain photometric 
measurements even under non-photometric conditions.

Unfortunately the atmospheric conditions were not stable
during the observations and no simultaneous photometry was available
which could have been used to correct for the
atmospheric variations.
In addition a careful check of the recorded standard star spectra showed
that a significant loss of light occured at the spectrograph slit.

\begin{figure*}
\vspace{0cm}
\mbox{
\includegraphics[clip,width=5cm,height=0.49\textwidth,angle=90]{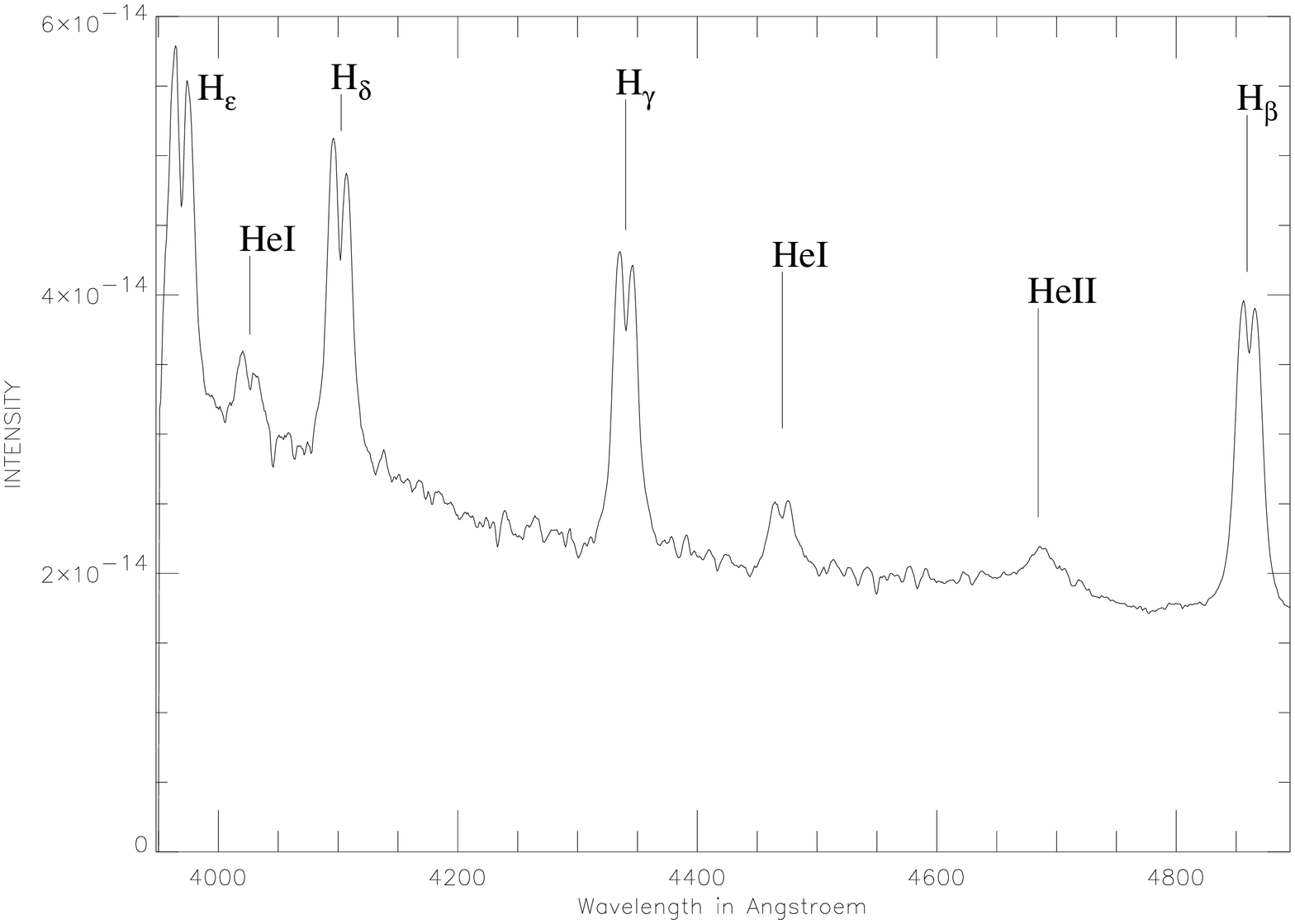}
\hspace{0cm}
\includegraphics[clip,width=5cm,height=0.49\textwidth,angle=90]{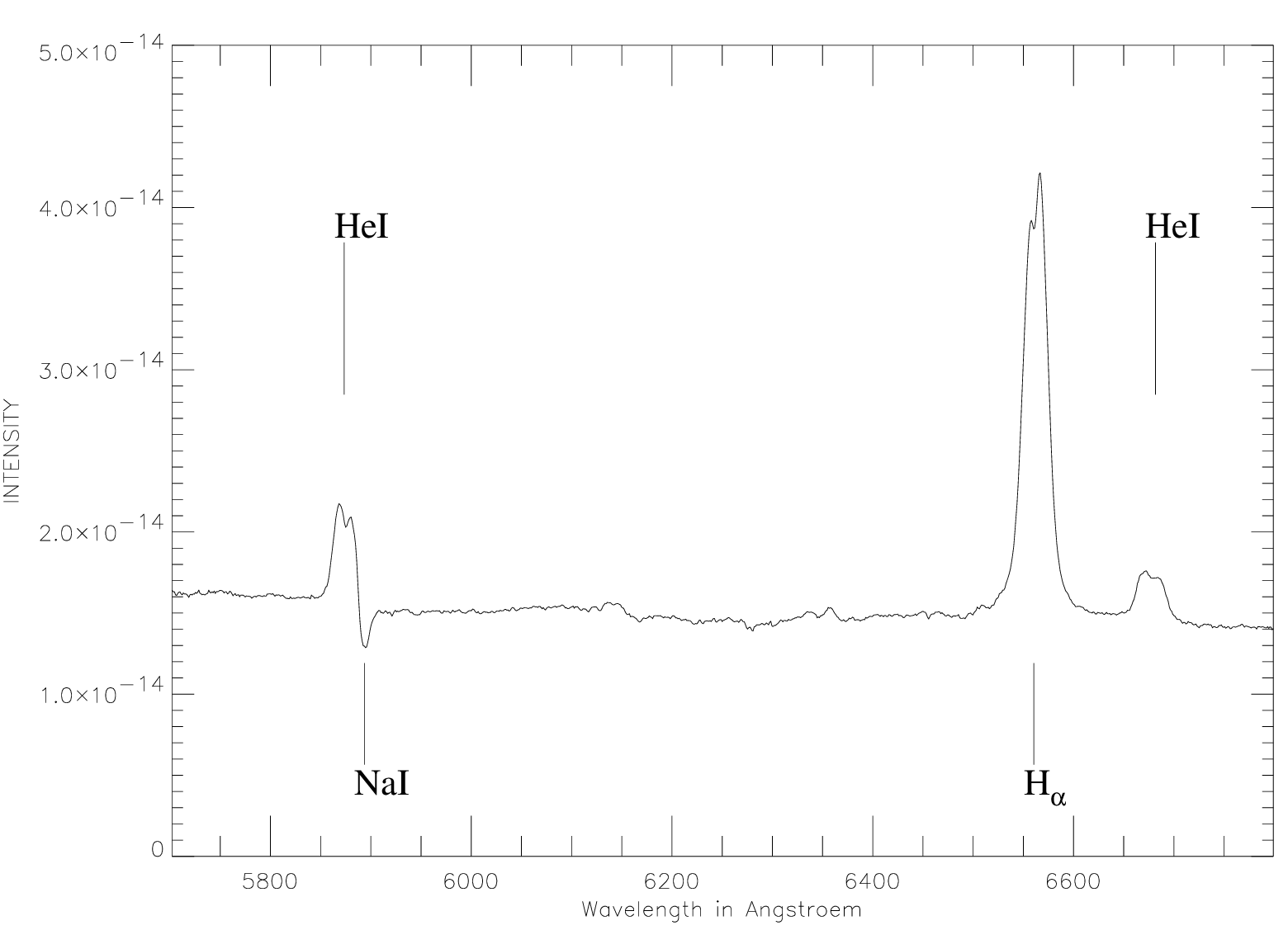}
\vspace{0cm}
}
\caption{\label{MeanQuSpec} \small Mean spectra of EX\,Dra observed during
quiescence. Before averaging the spectra were shifted according to
the radial velocity $K_1\sin(\phi) =  167\,\mbox{km}\,\mbox{s}^{-1}\sin(\phi)$.
Intensities in ergs\,s$^{-1}$\,cm$^{-2}$\,{\AA}$^{-1}$. The energy
flux is accurate to a constant factor, because of slit
loss of the flux star spectra. See the text for more details.
}
\end{figure*}

\begin{figure*}
\vspace{0cm}
\mbox{
\includegraphics[clip,width=5cm,height=0.49\textwidth,angle=90]
                {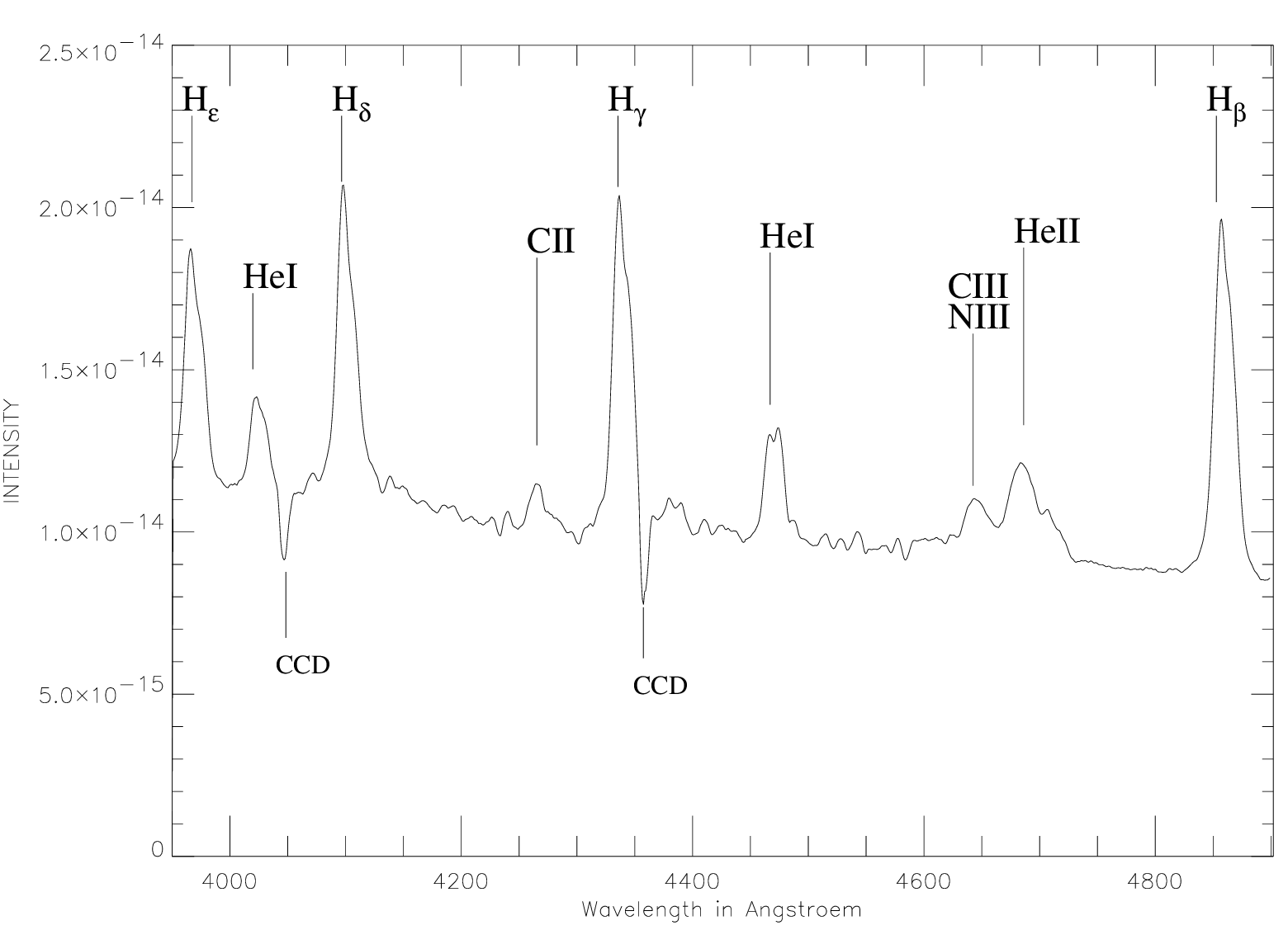}
\hspace{0cm}
\includegraphics[clip,width=5cm,height=0.49\textwidth,angle=90]
                {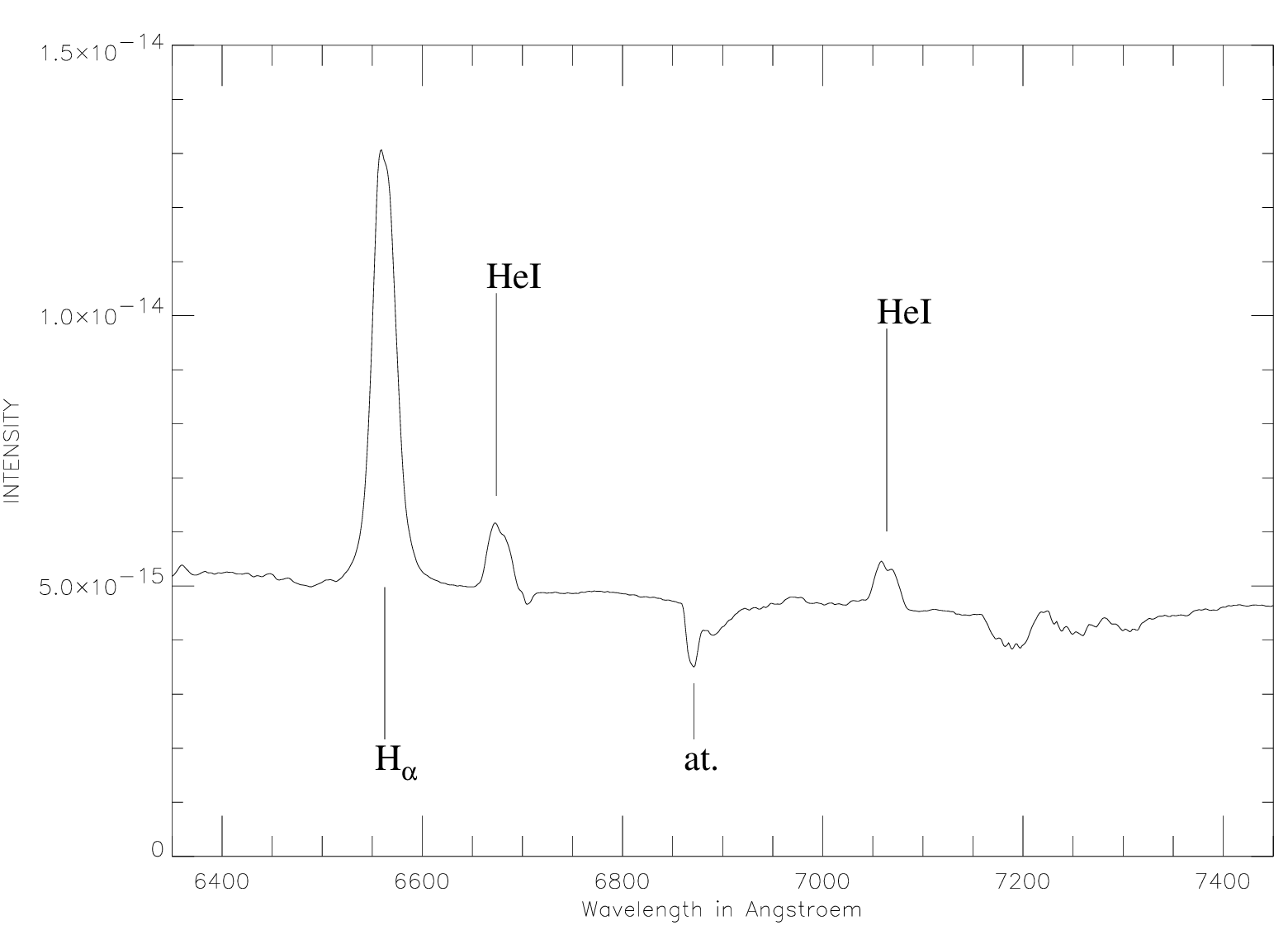}
\vspace{0cm}}
\caption{\label{MeanOutSpec}\small Mean spectra of EX\,Dra observed
during outburst. Before averaging the
spectra were shifted according to the radial velocity
$K_1\sin(\phi) =  167\,\mbox{km}\,\mbox{s}^{-1}\sin(\phi)$.
Intensities in ergs\,s$^{-1}$\,cm$^{-2}$\,{\AA}$^{-1}$. The flux is
accurate to a constant factor, because of slit
loss of the flux star spectra.
This fact could explain why the energy flux of the
system appears to be lower 
during outburst than during quiescence.
See the text for further discussion.
}
\end{figure*}

Therefore a different approach had to be made: the short term atmospheric
variations were corrected by means of \emph{mean} photometric B and R
light curves taken during the quiescent and the outburst states of EX\,Dra,
whereas the wavelength--dependent sensitivity of the atmosphere and
the instrument were corrected by using the recorded spectra of the
flux stars BD\,+28$^\circ$4211 (1992) and Wolf\,1346 (1993). 
A significant number of photons is lost when a narrow spectrograph
slit is used for both the object as well as the standard star spectra. The
lack of standard star spectra recorded with a wide slit prevented
us from calculating the so--called 'slit loss'. Therefore
the resulting spectral flux distribution of
the object spectra is accurate to a constant factor.
%


\section{Analysis and results}
\label{analyses}

\subsection{Mean quiescence spectra}

Fig.~\ref{MeanQuSpec} presents the
mean spectra of EX\,Dra	in quiescence, which are based on
69 single spectra in the blue
and 68 in the red wavelength range. Before averaging the single spectra
were shifted according to 
the radial velocity
$K_1\sin(\phi)$, with $K_1 = 167~\mbox{km}\,\mbox{s}^{-1}$ 
(\cite{Fiedler97}). 
The spectra show strong, broad emission lines of hydrogen with
equivalent widths up to 55\,{\AA} for H$_{\alpha}$. Neutral helium is present
in emission in both the singlet line at $\lambda$\,6678\,{\AA} and the
triplet lines at $\lambda\lambda$\,4471, 4026, 5876\,{\AA}. 
\ion{He}{ii} is weakly present at $\lambda$\,4686\,{\AA}. 
Lines from other (highly) ionized species are rarely seen during
quiescence, except several \ion{Fe}{ii} 
($\lambda\lambda,$\,4924, 5018, 5169\,{\AA}) lines.

The hydrogen and helium emission lines show the typical 
double--peaked profiles associated with accretion disks and are therefore
mainly formed within the disk. These profiles are superimposed
by asymmetric structures produced by anisotropically radiating 
emission sites of the system. The Balmer lines H$_{\beta}$,
H$_{\gamma}$, H$_{\delta}$, H$_{\epsilon}$ and the \ion{He}{i}
($\lambda$\,6678\,{\AA}) line show an asymmetry in form of an enhanced
\emph{blue--shifted} peak in the averaged spectra. 
The phase--resolved single spectra of these lines show clearly
that the maximum of line emission is seen during the phases
before the eclipse ($\phi \approx 0.7\dots0.95$) in the blue--shifted
part of the line profiles.
Obviously it can be attributed to
the interaction between the infalling gas stream and the outer parts of
the disk which produces an anisotropically radiating region of 
enhanced emission.

Departing from that canonical picture the
line profile of the H$_{\alpha}$ line shows an asymmetry in form
of an enhanced \emph{red--shifted} peak. 
Obviously at lower temperatures
the emission of the bright spot is exceeded by 
other sources of emission within the system. 
The phase--resolved line profiles reflect a complex structure
indicating that several emission sites contribute to the
H$_{\alpha}$ line flux, but do not allow a separation of the different
components. 
This will be further discussed in Sect.~\ref{DopplerImaging} in the
context of the Doppler maps.

The secondary star is visible in several \ion{Ca}{i} absorption
lines in the red part of the optical wavelength range. A detailed
\mbox{analysis} of these lines can be found in 
Fiedler~et~al. (1997). 


\subsection{Mean outburst spectra}

The mean spectra of EX\,Dra in outburst are displayed in
Fig.~\ref{MeanOutSpec} as the average of 16 single spectra for the blue
and 16 for the red spectral range.
As discussed in Sect.~\ref{Calibration} the spectra of the spectrophotometric
standard stars suffered from significant slit loss and therefore the
computed flux
of the spectra of EX\,Dra is accurate to a constant factor. 
The energy flux of the system 
appears to be higher in quiescence than
outburst.
The mentioned slit loss should account for this contradiction, 
since we know from photometric
measurements taken the night 
after the spectroscopic observations in 1993 that the optical flux
was significantly enhanced compared to July 1992.

The most remarkable features in the outburst spectra are
dominant emission lines of the Balmer series and of neutral helium,
whereas outburst spectra  generally show broad Balmer absorption lines
with weak emission cores.  
A similar behaviour was also seen in IP\,Peg: 
Marsh \& Horne (1990)
recorded outburst spectra of IP\,Peg and observed
enhanced emission line fluxes of all Balmer and helium lines with
\ion{He}{ii} ($\lambda$\,4686\,{\AA}) becoming the strongest line at visual
wavelengths. 
The systems \object{Z\,Cha} (\cite{Vogt82}) and
\object{OY\,Car} 
(\cite{laDous91}) also 
show emission lines during outburst. The high orbital 
inclination of these systems probably account for 
the strong emission lines during eruption, because the flux in the
optically thick continuum of the accretion disk is at large
inclinations reduced by projection and limb darkening in favor of
emission lines formed above the disk.

The profiles of the Balmer lines are almost
single--peaked in the mean spectra of EX\,Dra in outburst, although
there are still some small indications of double--peaked structures visible.
As a remarkable exception the \ion{He}{i} ($\lambda$\,4471\,{\AA})
line clearly displays a double--peaked profile in the mean outburst spectra.

High--excitation lines which are only weakly or not at all present during
quiescence emerge during outburst, like \ion{He}{ii} ($\lambda$\,4686\,{\AA}),
\ion{C}{ii} ($\lambda$\,4267\,{\AA}) and the \ion{C}{iii}/\ion{N}{iii}
($\lambda\lambda$\,4634\dots4651\,{\AA}) blend.

\ion{He}{ii} emission is often a hint for a strong magnetic field of
the white dwarf, but there is no evidence for magnetic accretion in
EX\,Dra. 
Patterson \& Raymond (1985)
show that \ion{He}{ii}
($\lambda$\,4686\,{\AA}) emission can be produced in the upper layer of
the disk by reprocessing soft X--rays from the boundary layer 
when the mass transfer rate $\dot{\mbox{M}}(\mbox{d})$ exceeds
10$^{-9}\,\mbox{M}_{\sun}/\mbox{y}$. 
Under the premise that $\dot{\mbox{M}}(\mbox{d})$ in EX\,Dra in the
recorded outburst state satis\-fies this condition 
the detected \ion{He}{ii} outburst line 
could be formed in the (upper layer of the) disk by recombination
following photoionization by the boundary layer.
\ion{He}{ii} is blended by \ion{C}{iii}/\ion{N}{iii} emission, which
prevents us from Doppler imaging this line. The time--resolved line
profiles 
show that the line forming region can not be too
extended since most of the \ion{He}{ii} emission is eclipsed at 
phase $\phi = 0$
by the secondary star.


\subsection{Doppler imaging in the quiescent state} 
\label{DopplerImaging}

Line profiles broadened by Doppler shifting retain an imprint of the
line emission region from which it originated. The Doppler tomography
is an imaging technique, developed by 
Marsh \& Horne (1988), 
which makes use
of the close relationship between the observed emission line flux and
the velocity profiles to obtain the distribution of line emission over
the surface of the disk in velocity space. 

Under the premise that Doppler shifting is the only broadening
mechanism of significance it is possible to locate the line forming
regions of the binary system by means of Doppler tomography.

The Doppler broadened line profiles represent a projection of the
velocity distribution in the direction of the observer's line of sight, while
rotation of the binary gives the observer a continuously
varying sequence of velocity projections. This combination of Doppler
shifting and binary rotation provides sufficient information for the
assembly of two--dimensional maps in velocity space ($V_x,V_y$).

While the observed line profiles are a projection of the velocity
distribution, a
back--projection algorithm applied to the observed data 
yields the emission distribution in velocity space.

The computation of a Doppler tomogram is based  upon the fundamental
fact, that a spot of emission in the binary system traces an 'S--wave' in the 
phase--sorted ('phase--folded') spectra, most evidently displayed by the 
features of the bright spots in many cataclysmic variables.
The coordinate system is defined by the 
X--axis pointing from the
white dwarf to the secondary star and the Y--axis in the
direction of motion of the secondary star.
The sinusoidal 'S--wave' radial velocity curve $V_R$ described by an emission
spot with velocity coordinates ($V_x,V_y$) is a function of the
orbital phase $\phi$:

\begin{equation}
\label{swave}
V_R(\phi,V_x,V_y) = \gamma - V_x\cos(2\pi\phi) +  V_y\sin(2\pi\phi)
\quad ,
\end{equation}
%
where $\gamma$ is the velocity of the center of gravity of the system.
Every pixel ($V_x,V_y$) of the Doppler map corresponds to
such a particular S--wave $V_R(\phi,V_x,V_y)$ in the
data.
By summing up the flux in the data along this S--wave
the mean flux of this particular Doppler map pixel is obtained.

To reconstruct Doppler maps for the emission lines of EX\,Dra a linear
tomography algorithm, the Fourier--Filtered--Back--projection (FFBP) is
used, this procedure is described in detail for example by
Horne (1991).
%
%
\begin{figure*}
\hbox{\mbox{}
\hfill\includegraphics[bbllx=14,bblly=26pt,bburx=525,bbury=720,clip,width=4cm]
{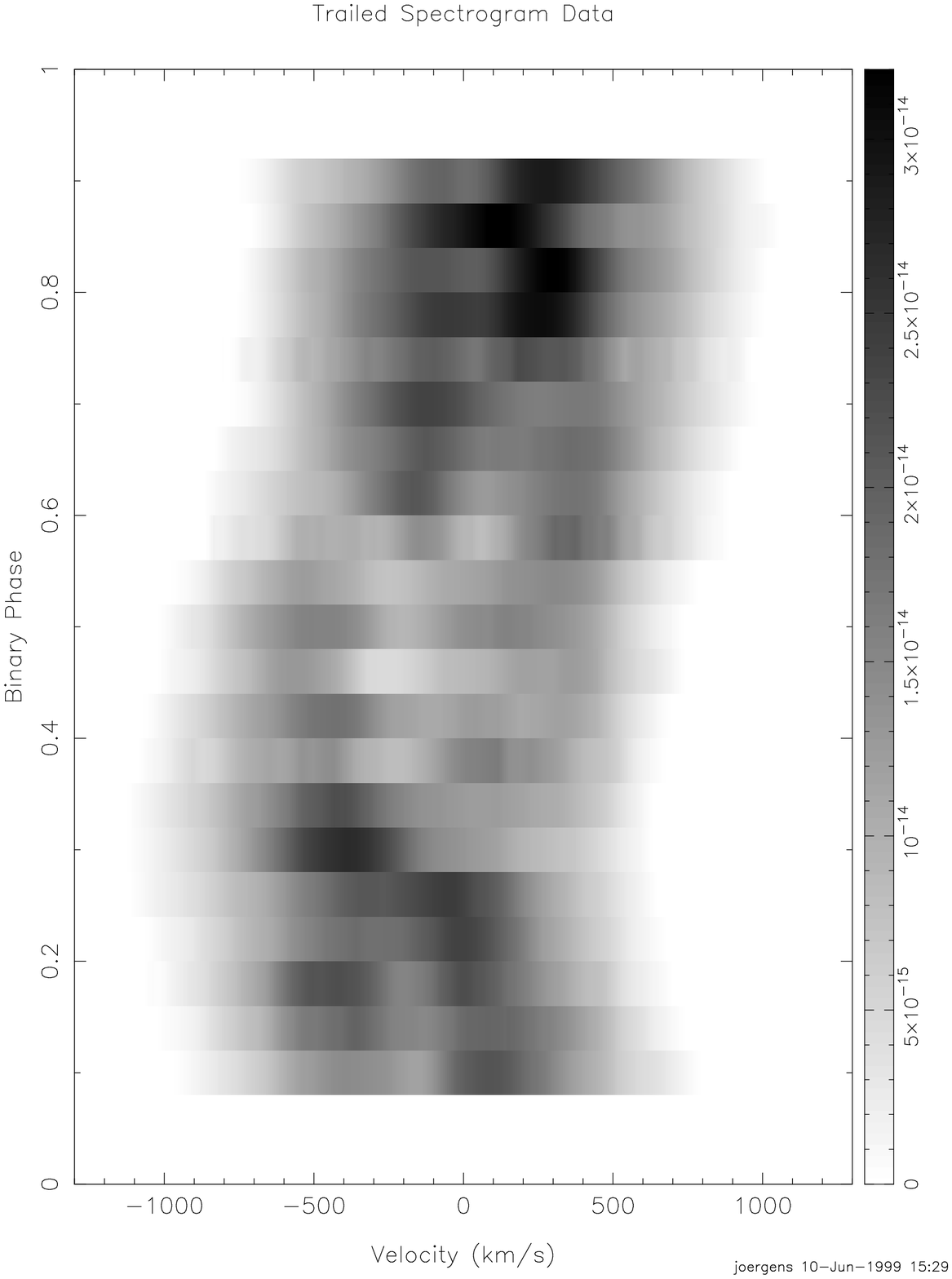}
\hfill\includegraphics[bbllx=14,bblly=26pt,bburx=525,bbury=720,clip,width=4cm]
{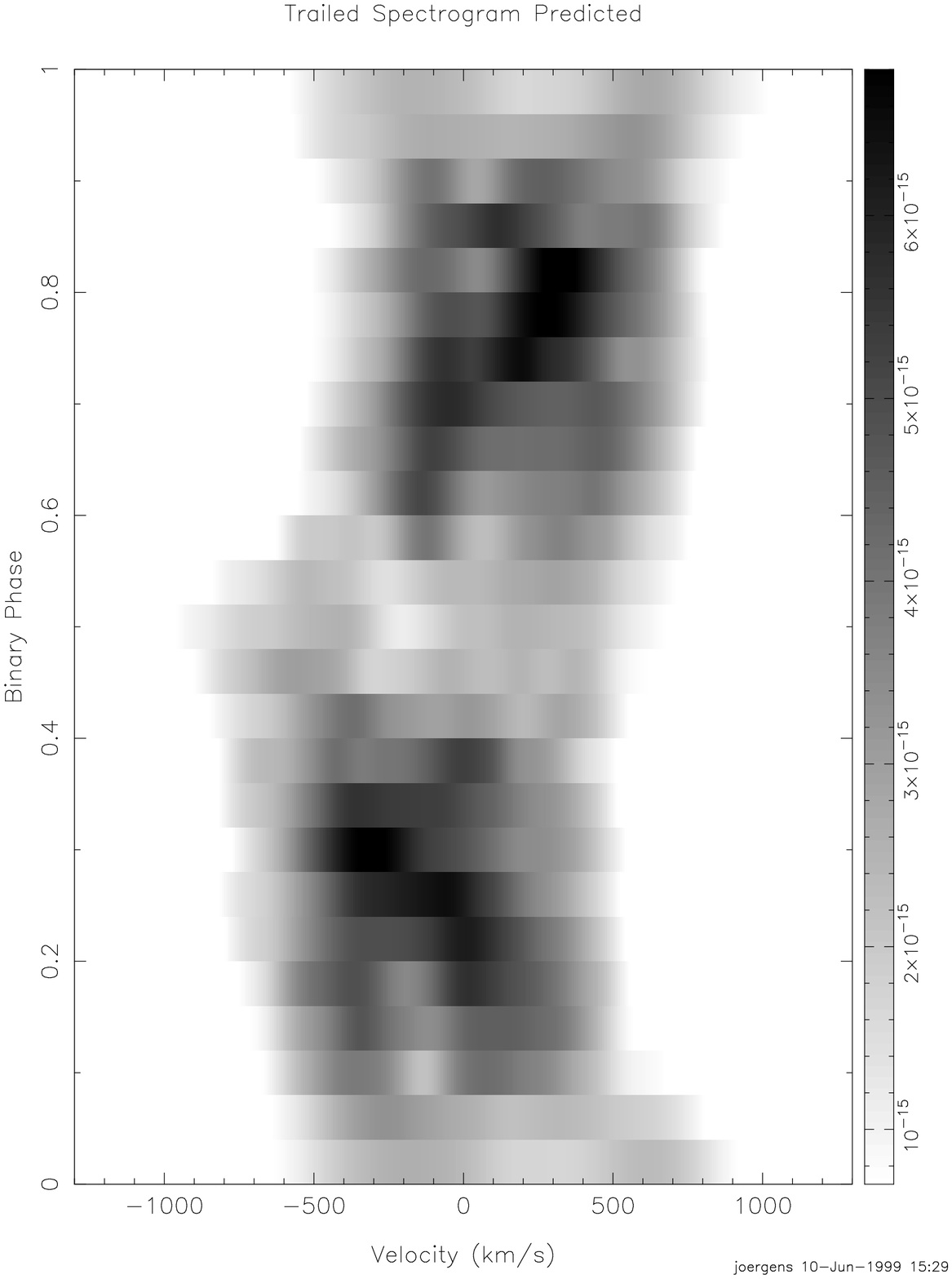}
\hfill\includegraphics[clip,width=5.2cm]{dm6563_6.ps.new}\hfill\mbox{}
}
\vspace{0.5cm}
\hbox{\mbox{}
\hfill\includegraphics[bbllx=14,bblly=26pt,bburx=525,bbury=720,clip,width=4cm]{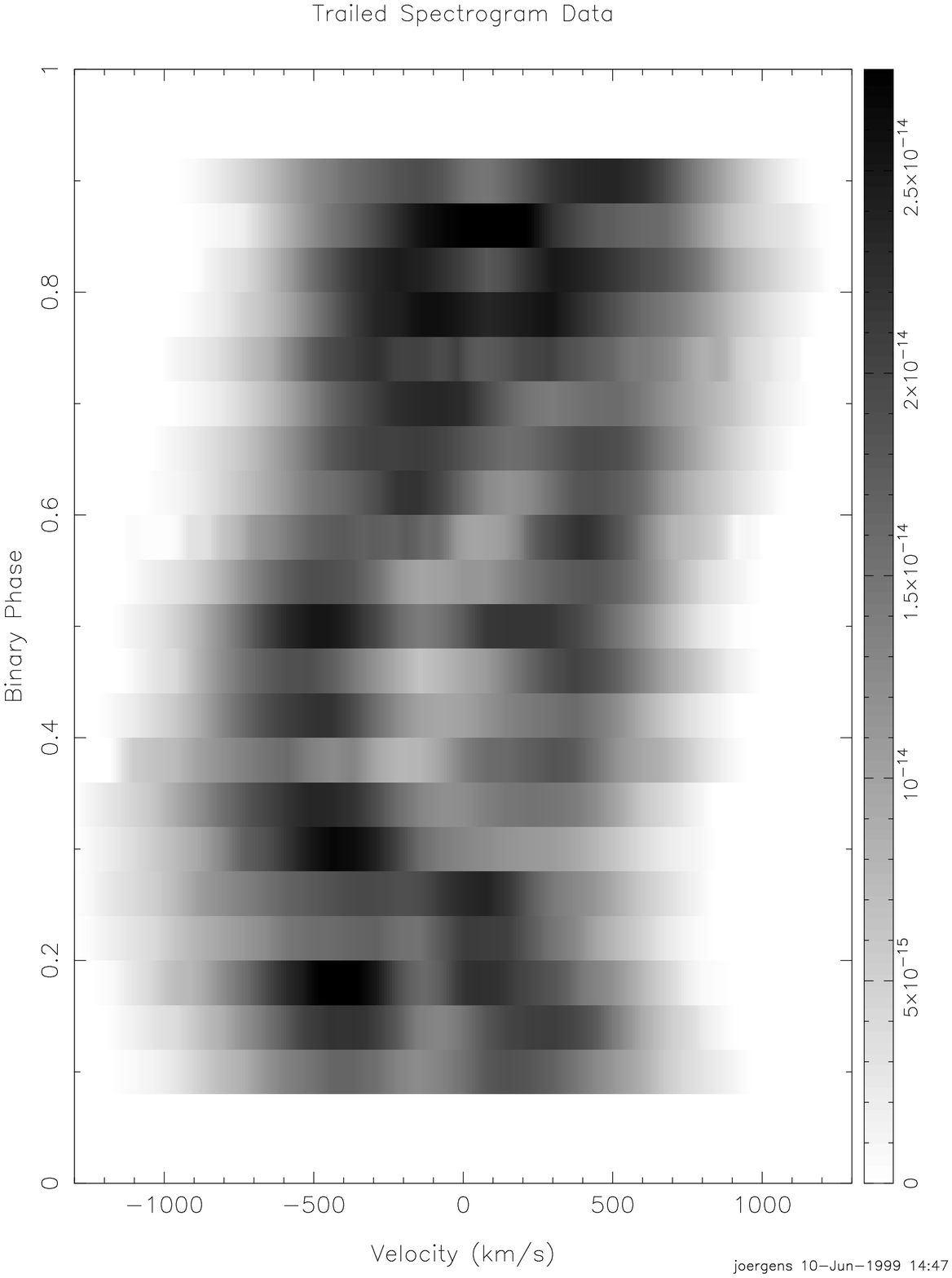}
\hfill\includegraphics[bbllx=14,bblly=26pt,bburx=525,bbury=720,clip,width=4cm]{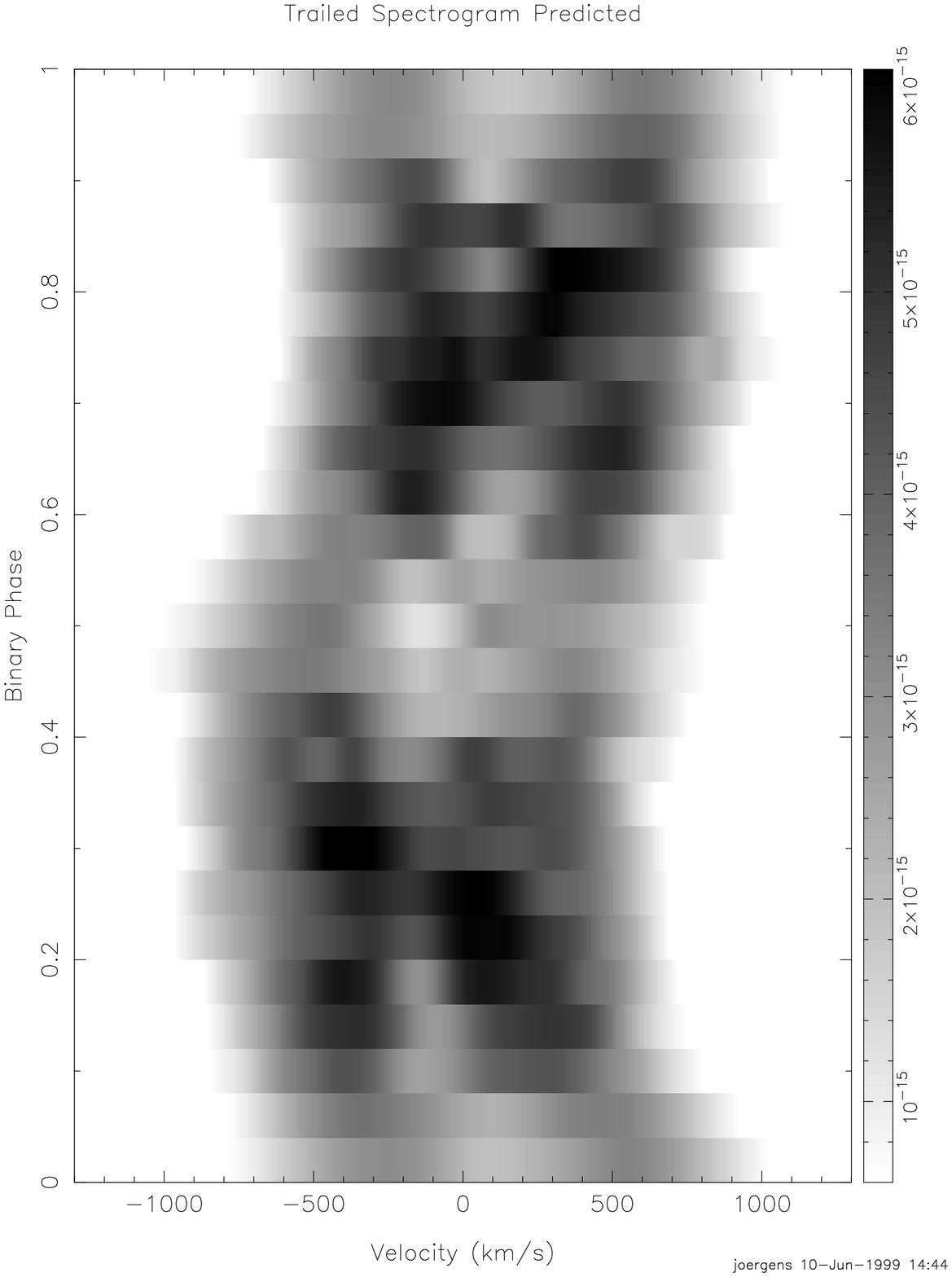}
\mbox{\includegraphics[clip,width=5.2cm]{dm4860_1.ps.new}}\hfill\mbox{}
}
\vspace{0.5cm}
\hbox{\mbox{}
\hfill\includegraphics[bbllx=14,bblly=26pt,bburx=525,bbury=720,clip,width=4cm]{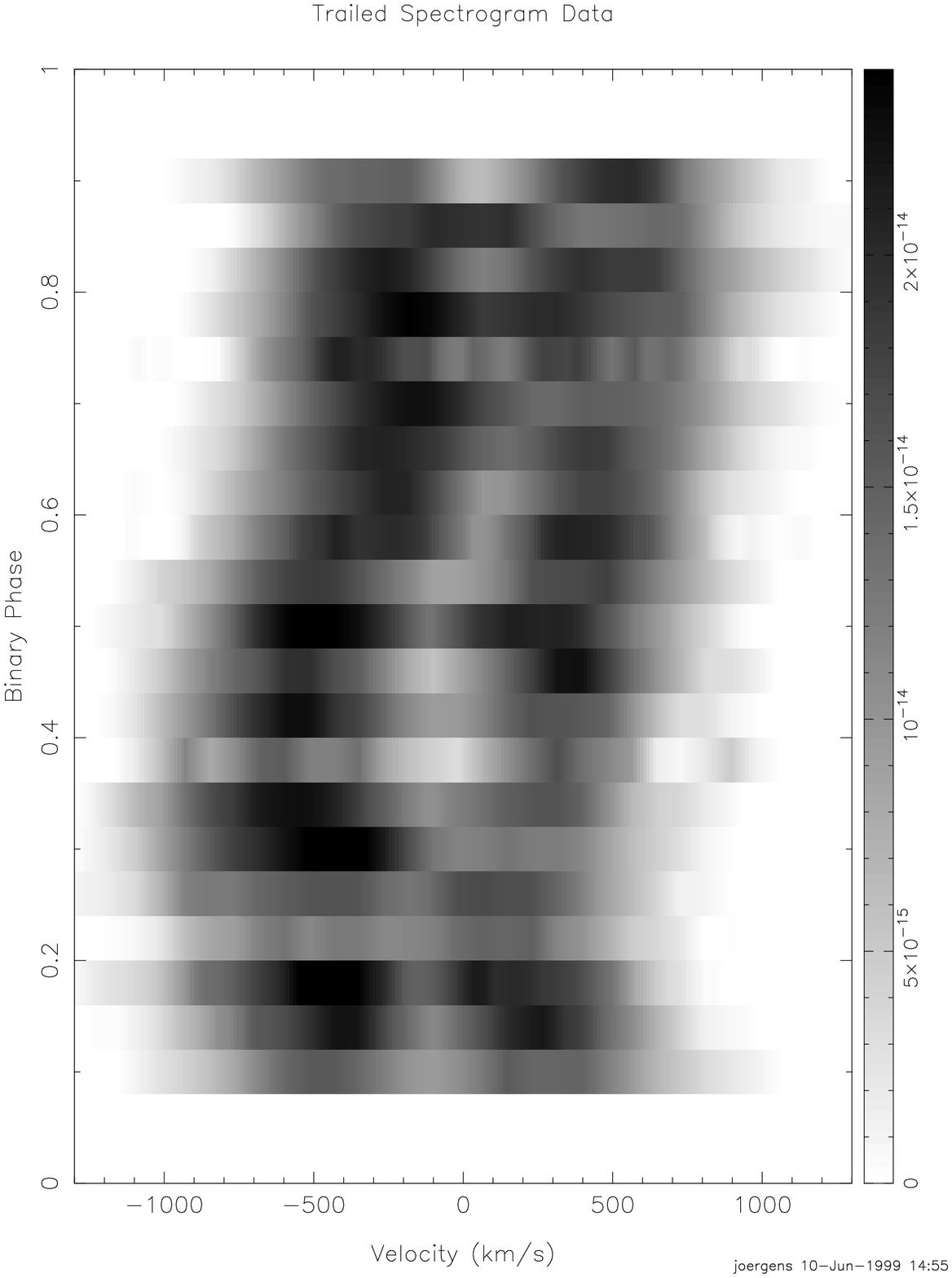}
\hfill\includegraphics[bbllx=14,bblly=26pt,bburx=525,bbury=720,clip,width=4cm]{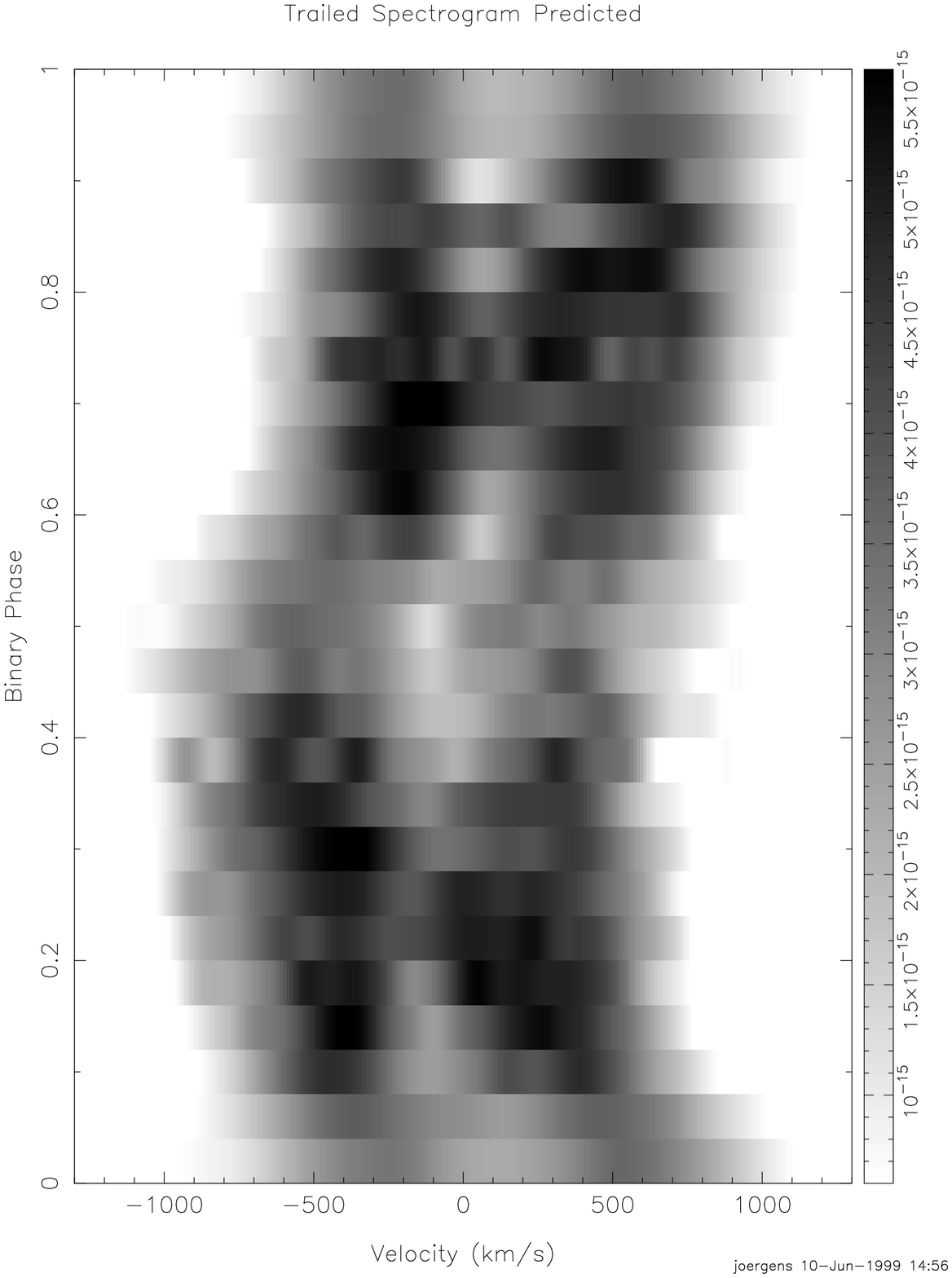}
\hfill\includegraphics[clip,width=5.2cm]{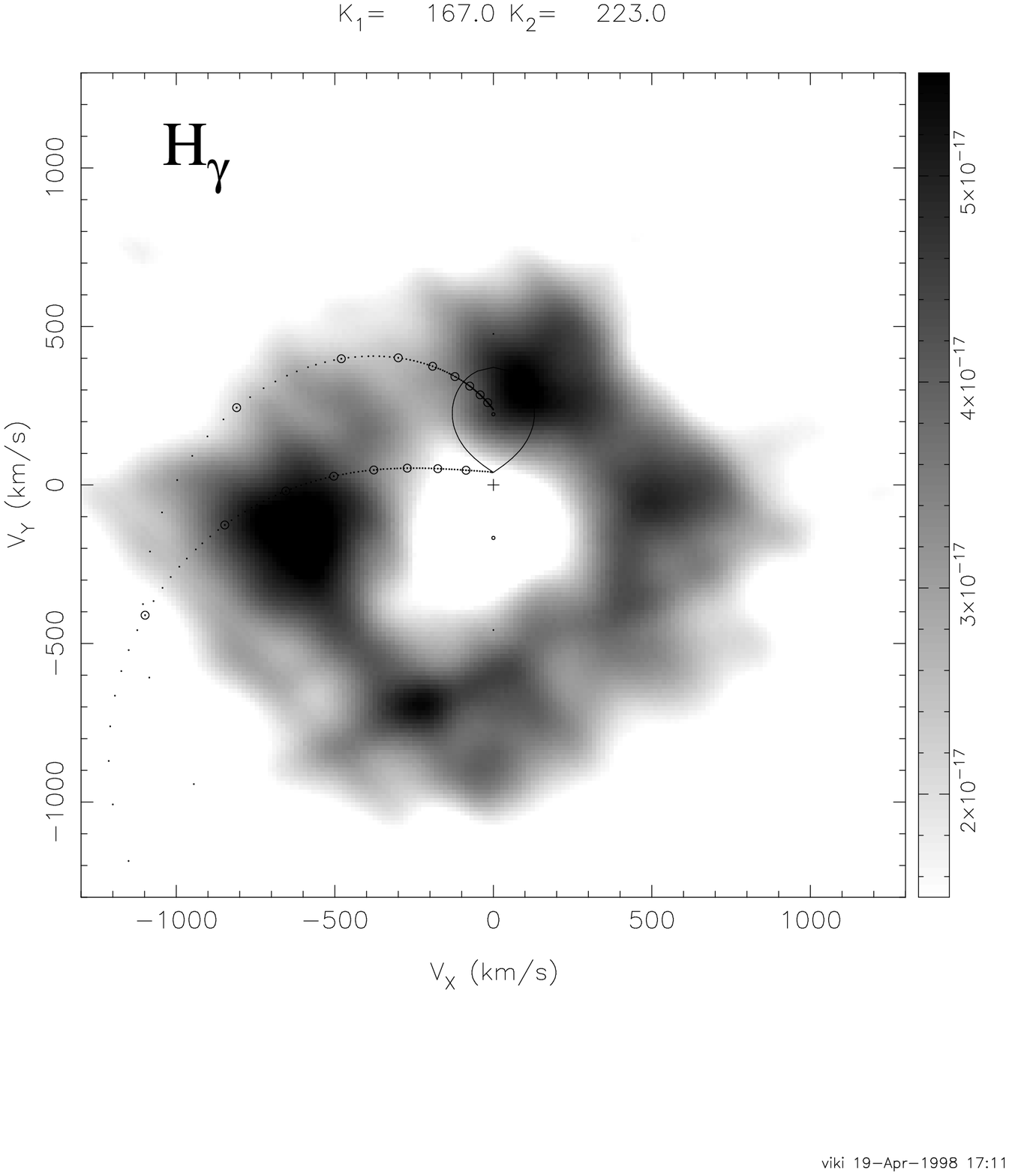}\hfill\mbox{}
}
\caption []{\label{quiescmap1} \small 
{
Doppler maps of H$_{\alpha}$, H$_{\beta}$ and H$_{\gamma}$ during the 
quiescent state of EX\,Dra (right column), the observed phase--folded spectra 
(left column) and the spectra reconstructed from the maps (middle).
The Doppler maps show a broad
ring--like structure and emission from the gas stream.
The map of H$_{\alpha}$ displays further a second
bright emission spot at V$_x \approx 0$\dots$+500$\,km\,s$^{-1}$ and
a faint hint of secondary emission.
} } 
\end{figure*}
\begin{figure*}
\vbox{
\vspace{0.5cm}
\hbox{\mbox{}
\hfill\includegraphics[bbllx=14,bblly=26pt,bburx=525,bbury=720,clip,width=4cm]{spec4341q.ps}
\hfill\includegraphics[bbllx=14,bblly=26pt,bburx=525,bbury=720,clip,width=4cm]{reco4341q.ps}
\hfill\includegraphics[clip,width=5.2cm]{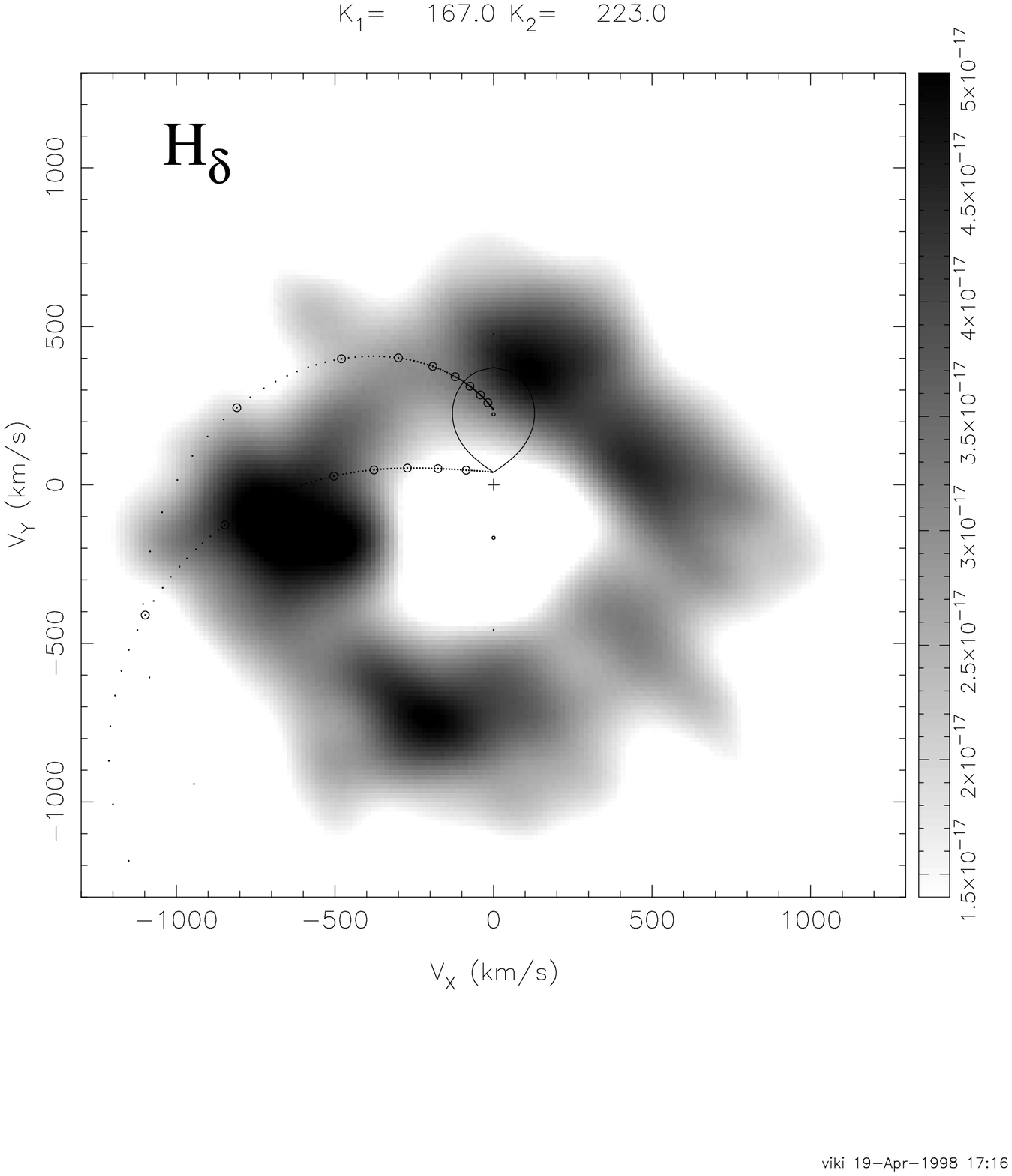}\hfill\mbox{}
}
\hbox{\mbox{}
\hfill\includegraphics[bbllx=14,bblly=26pt,bburx=525,bbury=720,clip,width=4cm]{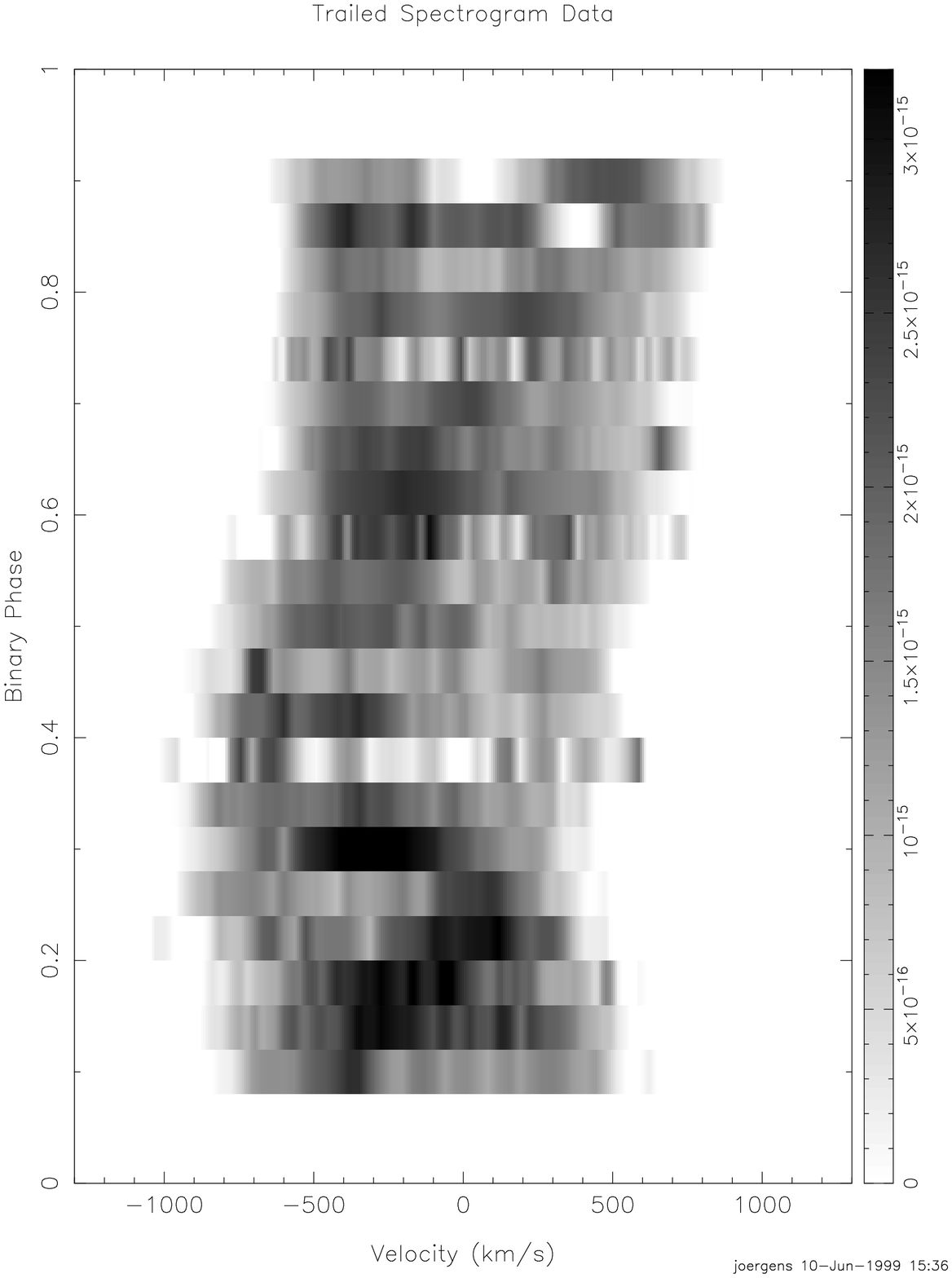}
\hfill\includegraphics[bbllx=14,bblly=26pt,bburx=525,bbury=720,clip,width=4cm]{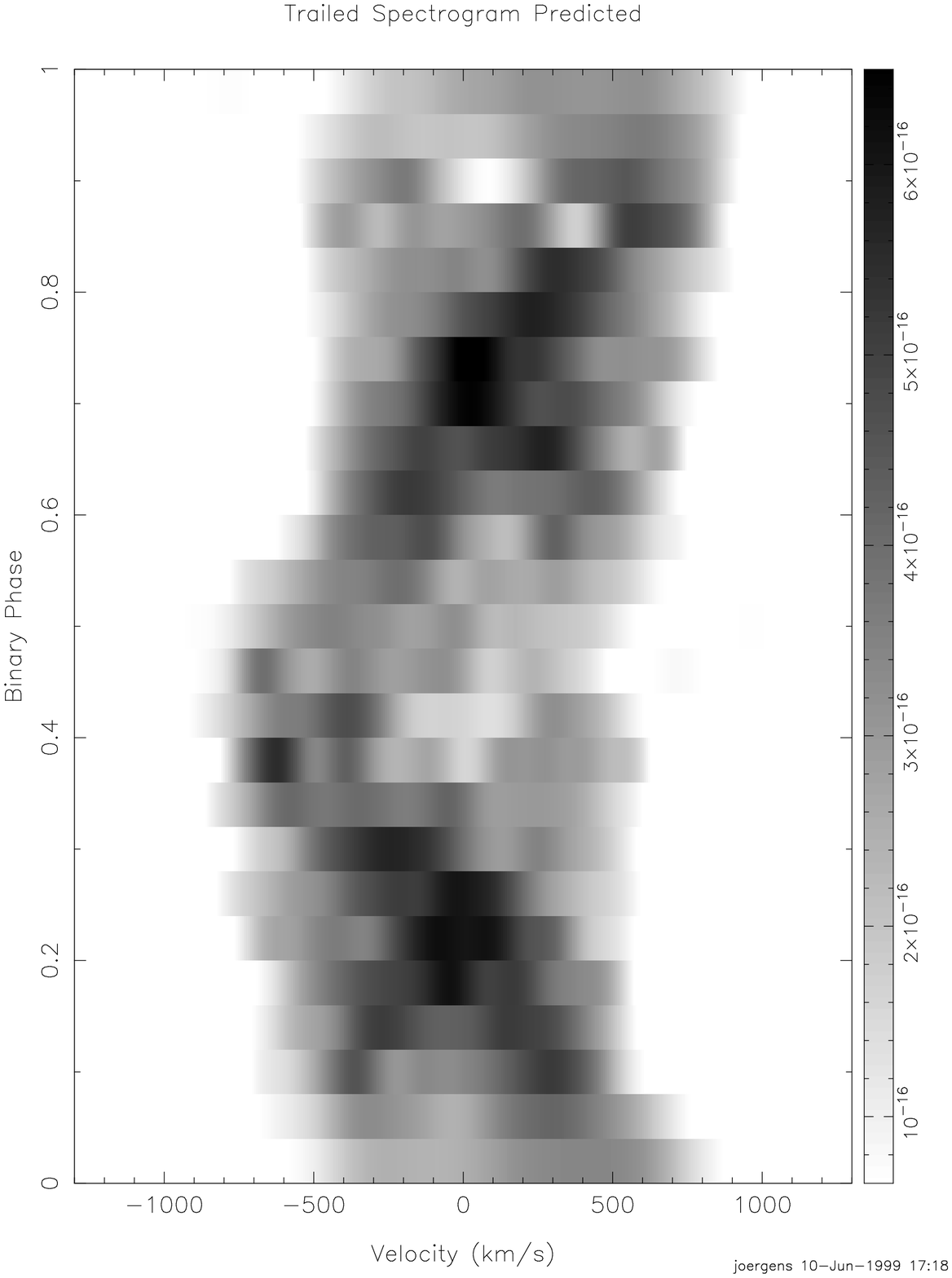}
\hfill\includegraphics[clip,width=5.2cm]{dm6678_1.ps.new}\hfill\mbox{}
}
\vspace{0.5cm}
\hbox{\mbox{}
\hfill\includegraphics[bbllx=14,bblly=26pt,bburx=525,bbury=720,clip,width=4cm]{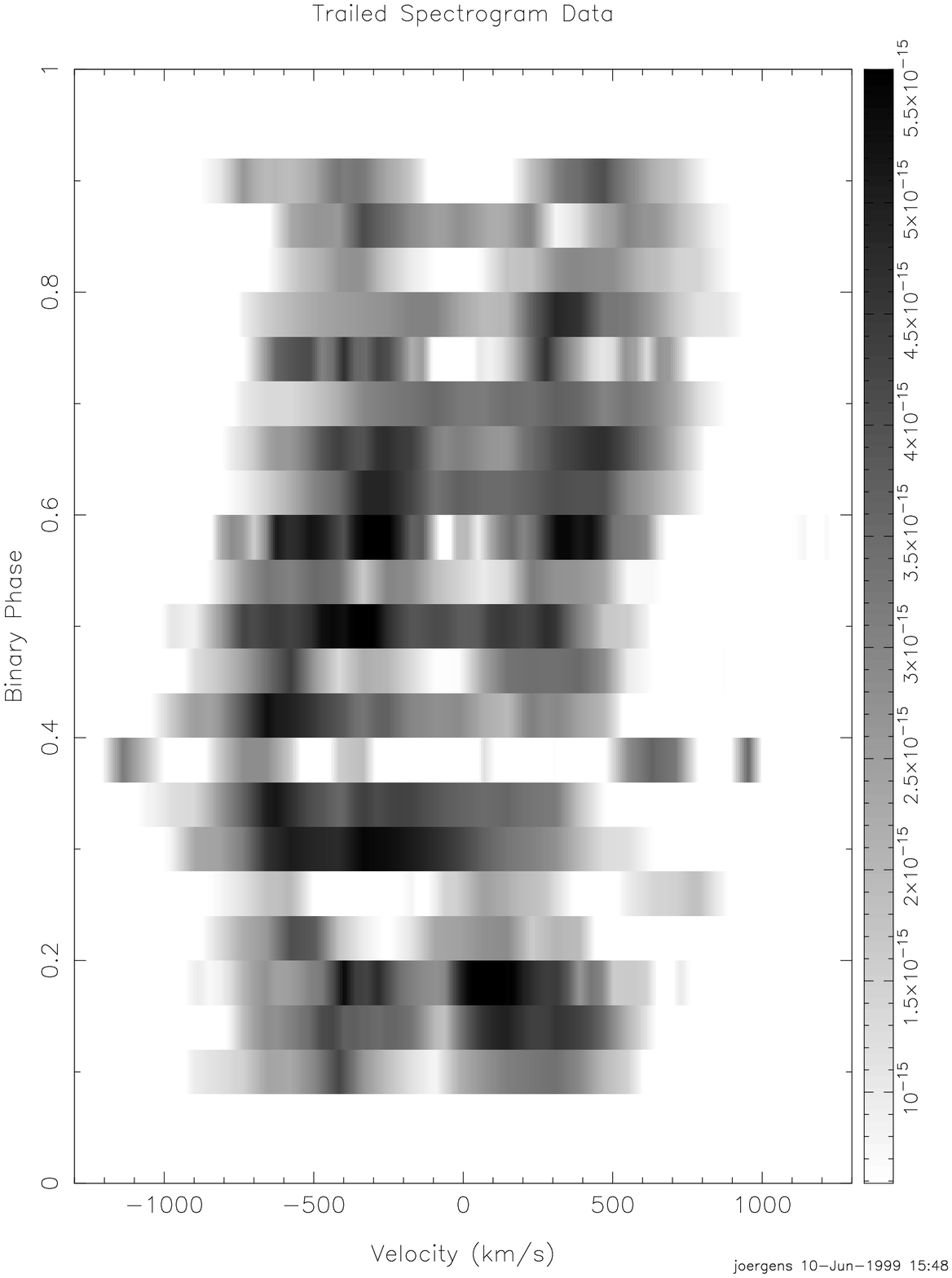}
\hfill\includegraphics[bbllx=14,bblly=26pt,bburx=525,bbury=720,clip,width=4cm]{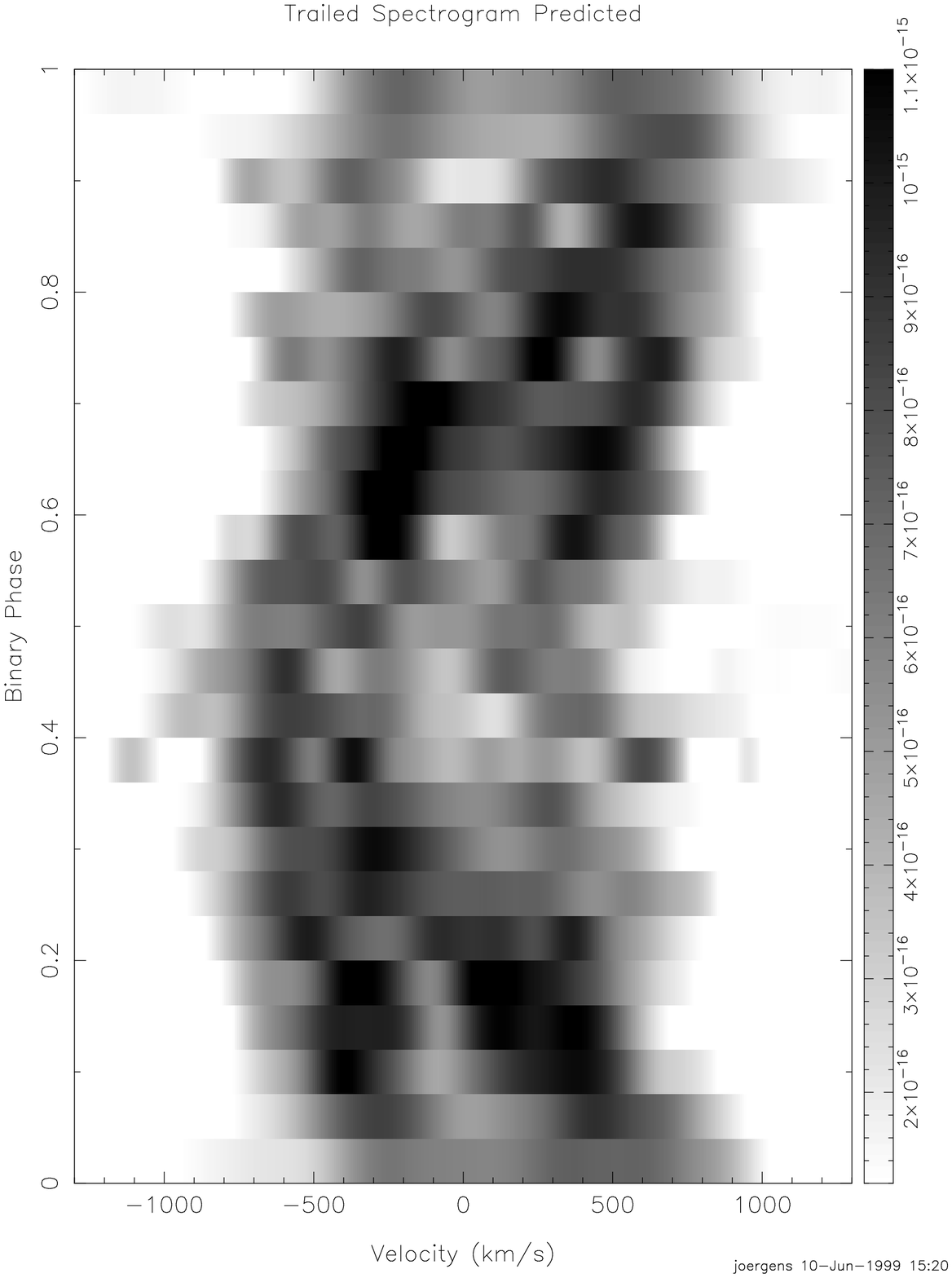}
\hfill\includegraphics[clip,width=5.2cm]{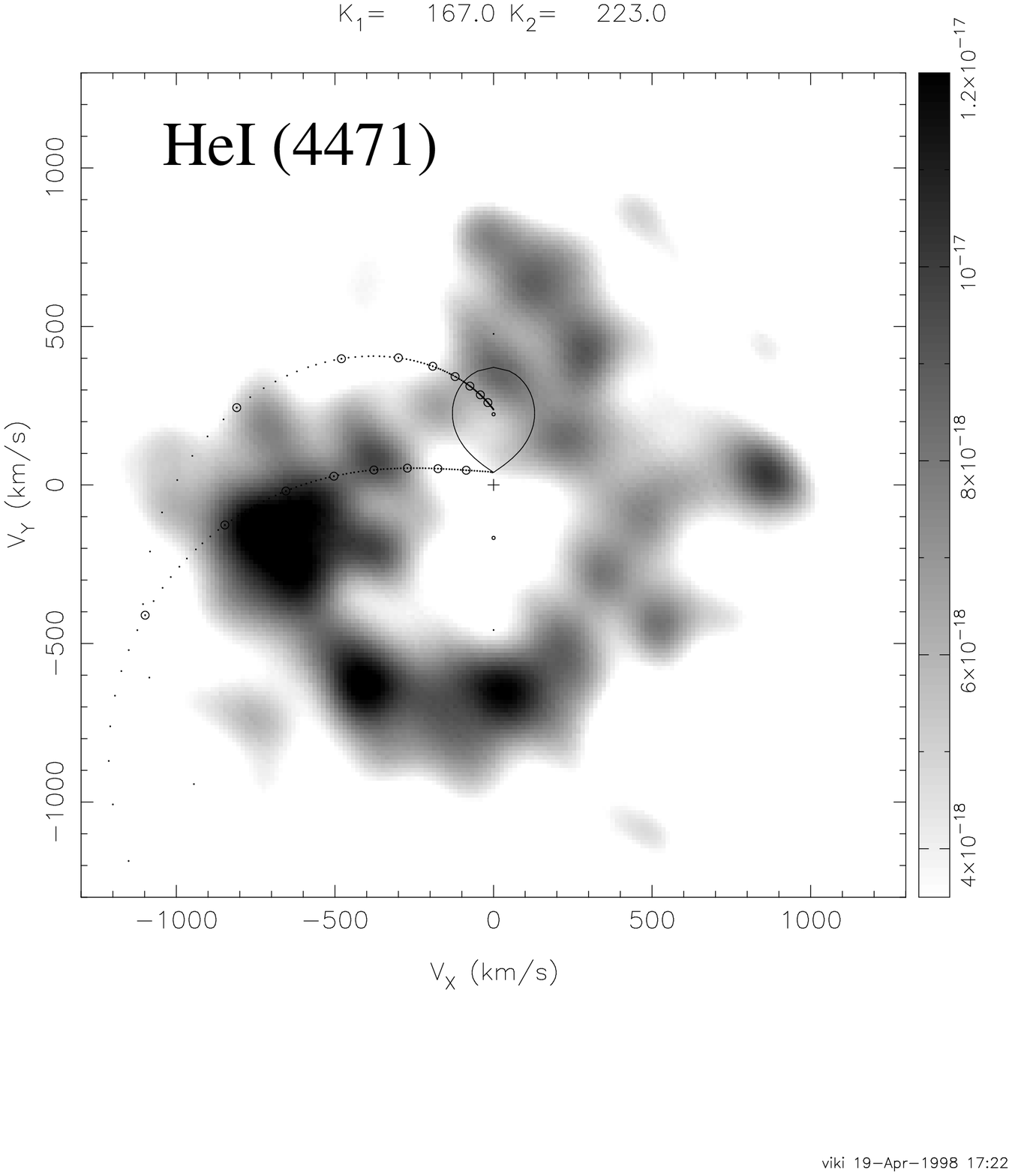}\hfill\mbox{}
}
\caption []{\label{quiescmap2}\small 
Doppler maps of  H$_{\delta}$ and 
\ion{He}{i} ($\lambda \lambda$ 6678 and 4471\,{\AA}) during the quiescent state
of EX\,Dra (right column), the observed phase--folded spectra 
(left column) and the spectra reconstructed from the maps (middle).
The H$_{\delta}$ map shows similar features as the H$_{\beta}$ and 
H$_{\gamma}$ map (cp. Fig.~\ref{quiescmap1}). The \ion{He}{i} maps display 
emission from the gas stream and the disk, but the data are quite noisy 
within this lines.
}
}
\end{figure*}
%
%
Since we do not assume any relation between
position and velocity coordinates during the eclipse,
spectra of the phases $\phi$ = 0.12$\dots$0.93
have not been taken into account. Before computing the tomograms the underlying
continuum flux has been subtracted from the individual emission lines.

The accretion disk in velocity space is turned inside out, since the
outer rim of the disk where the Keplerian velocity is smallest, maps
into the inner edge of the Doppler image.
The center of gravity of the system is by definition at 
the origin of
the Doppler map  and the white dwarf is represented by a point at
$(V_x,V_y) = (0,-\mbox{K}_1)$.   
Line forming regions of the secondary star map into a
Roche lobe--shaped region centered at $(V_x, V_y) = (0, \mbox{K}_2)$. 

A theoretical gas stream trajectory (lower arc in the Doppler maps)
based on K$_1$ and K$_2$ as given by 
Fiedler~et~al. (1997)
is 
plotted in the Doppler maps to help interpreting the reconstructed
emission distribution in velocity space. Starting 
with the velocity of the inner Lagrangian point L$_1$ the gas stream
trajectory describes an 
arc towards increased $-V_x$ and $-V_y$ velocities as the gas
particles are accelerated in the direction of the white dwarf ($-V_x$)
and approaching this central object spiral towards the center of gravity
($-V_y$).
Since the stream interacts with the disk, we may see enhanced
emission from the region of interaction
at velocities ranging between that of the
stream and the disk at that location.
Therefore a second trajectory (upper arc in the Doppler maps) is inserted, 
indicating the Keplerian velocity of the disk along the path of the
gas stream. 

\begin{figure*}
\vbox{
\hbox{\mbox{}
\hfill\includegraphics[bbllx=14,bblly=26pt,bburx=525,bbury=720,clip,width=4cm]{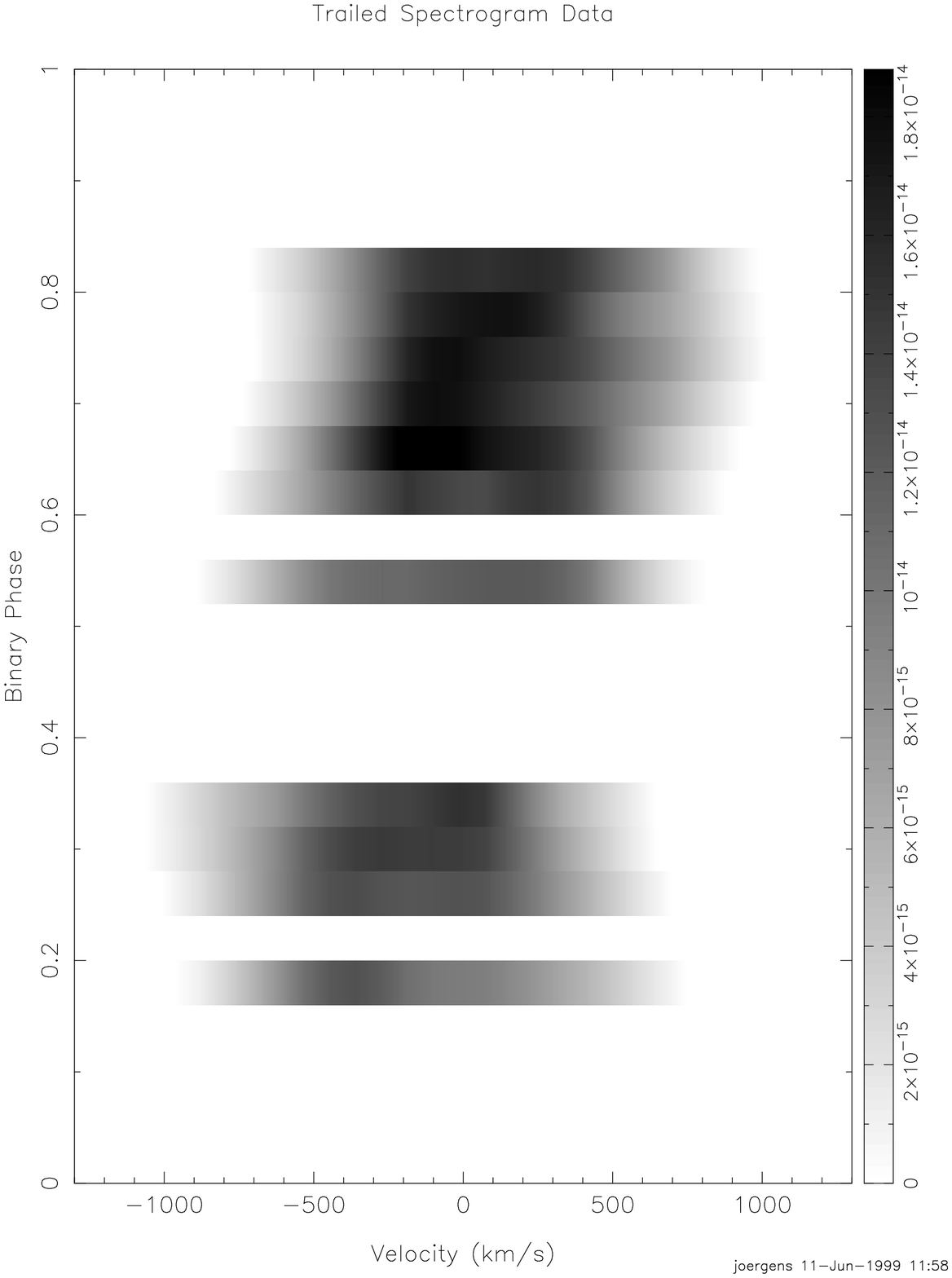}
\hfill\includegraphics[bbllx=14,bblly=26pt,bburx=525,bbury=720,clip,width=4cm]{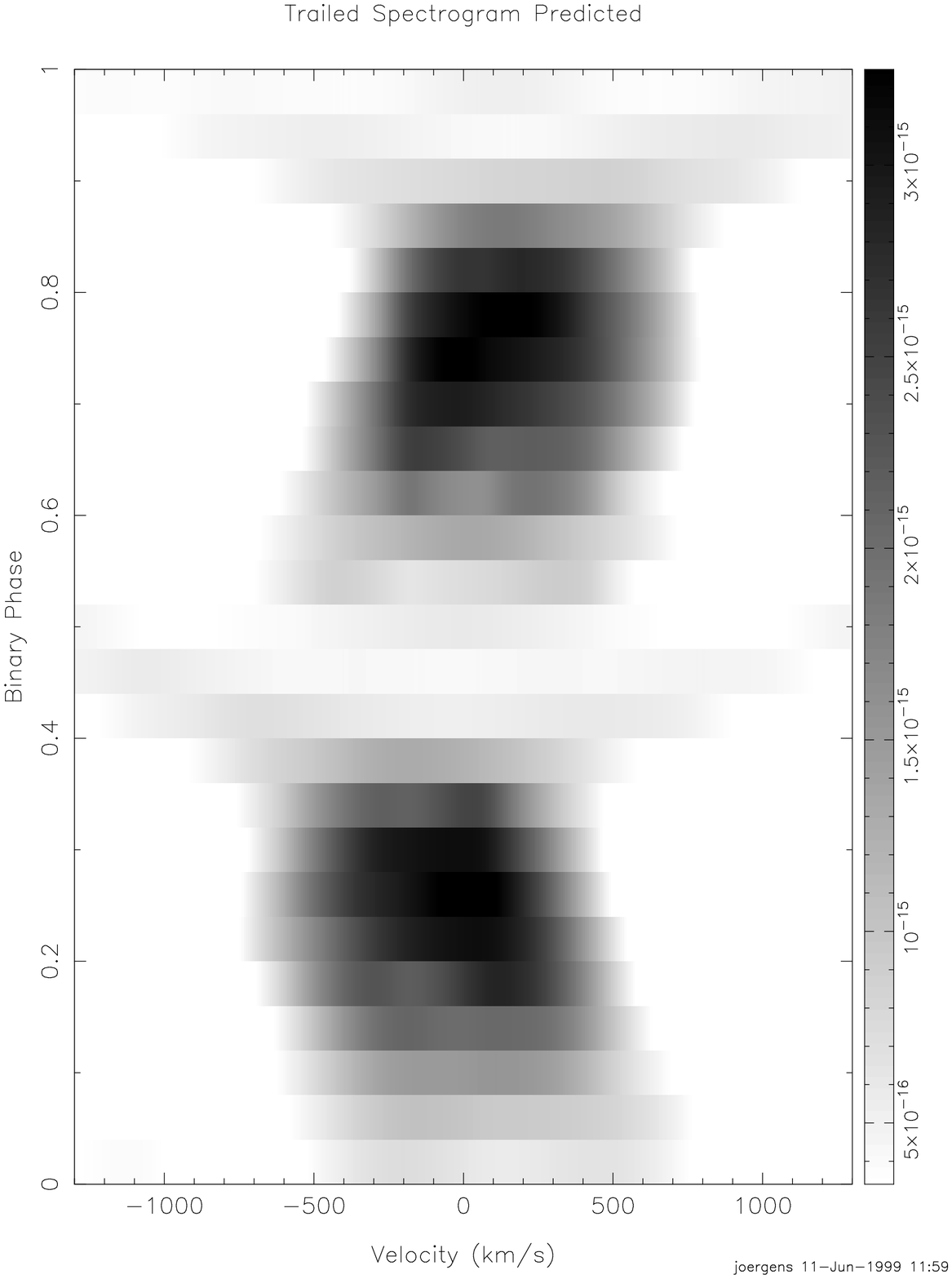}
\hfill\includegraphics[clip,width=5.2cm]{dm6563_ob1.ps.new}\hfill\mbox{}
}
\vspace{0.5cm}
\hbox{\mbox{}
\hfill\includegraphics[bbllx=14,bblly=26pt,bburx=525,bbury=720,clip,width=4cm]{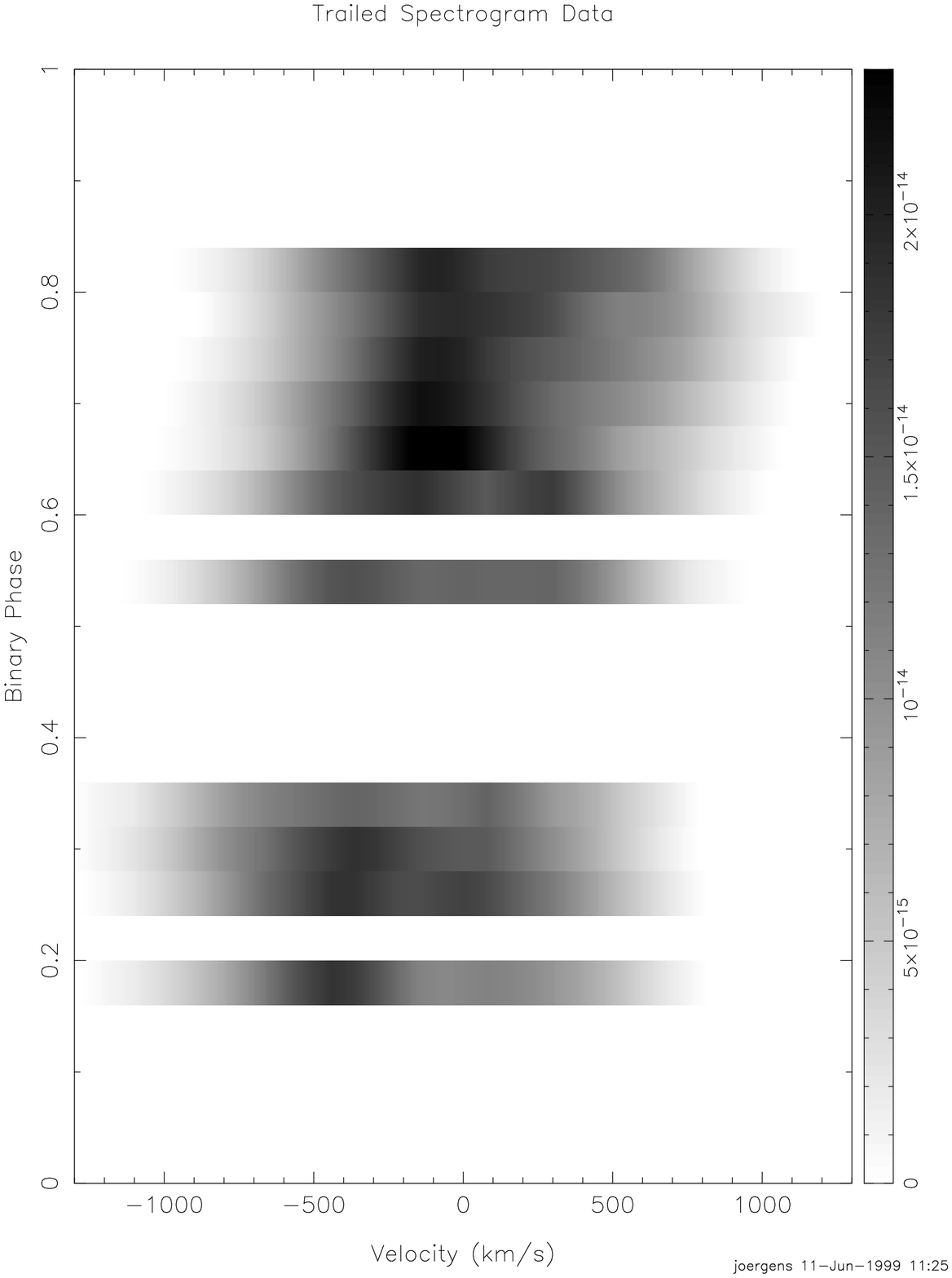}
\hfill\includegraphics[bbllx=14,bblly=26pt,bburx=525,bbury=720,clip,width=4cm]{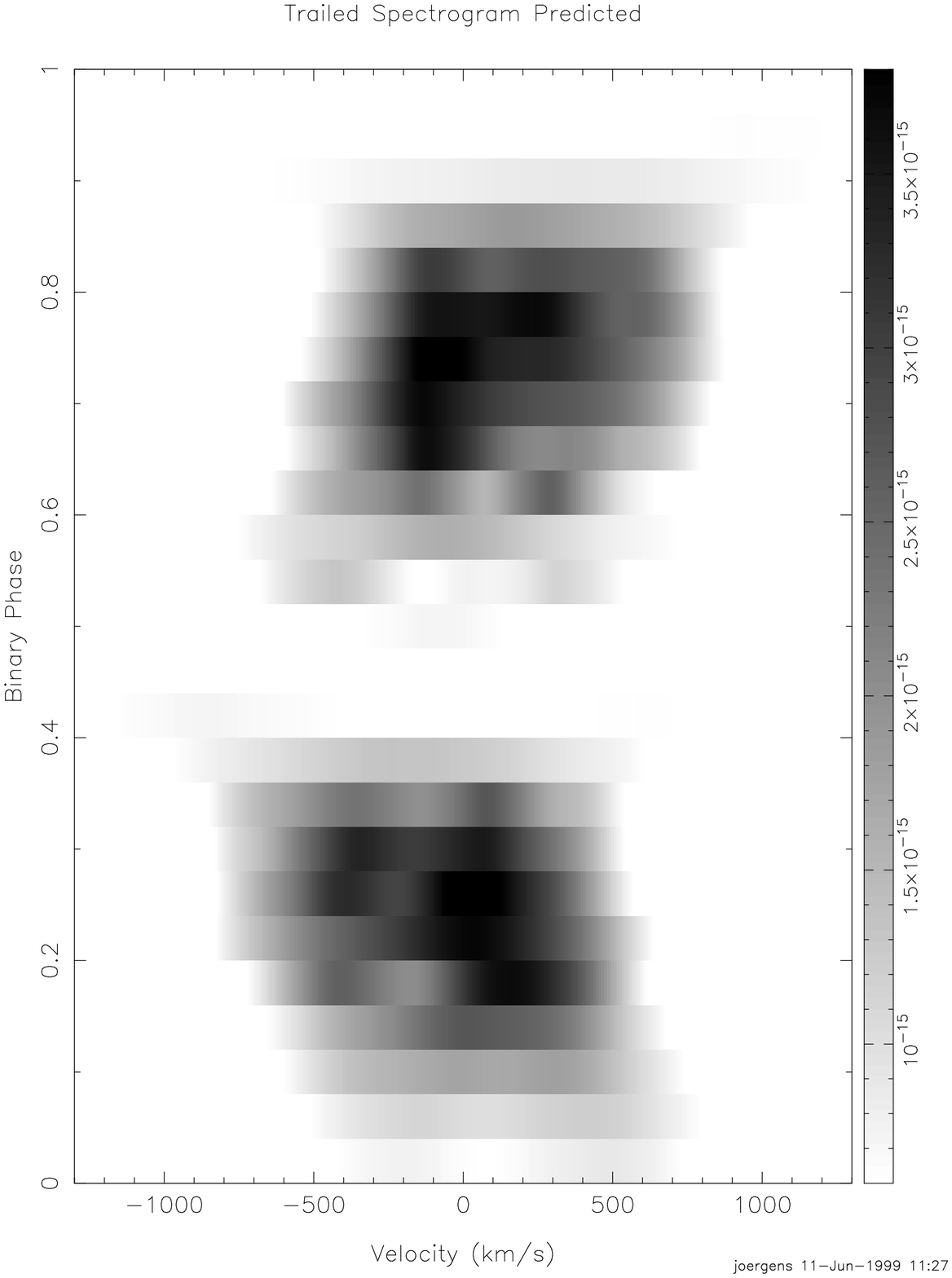}
\hfill\includegraphics[clip,width=5.2cm]{dm4861_ob1.ps.new}\hfill\mbox{}
}
\vspace{0.5cm}
\hbox{\mbox{}
\hfill\includegraphics[bbllx=14,bblly=26pt,bburx=525,bbury=720,clip,width=4cm]{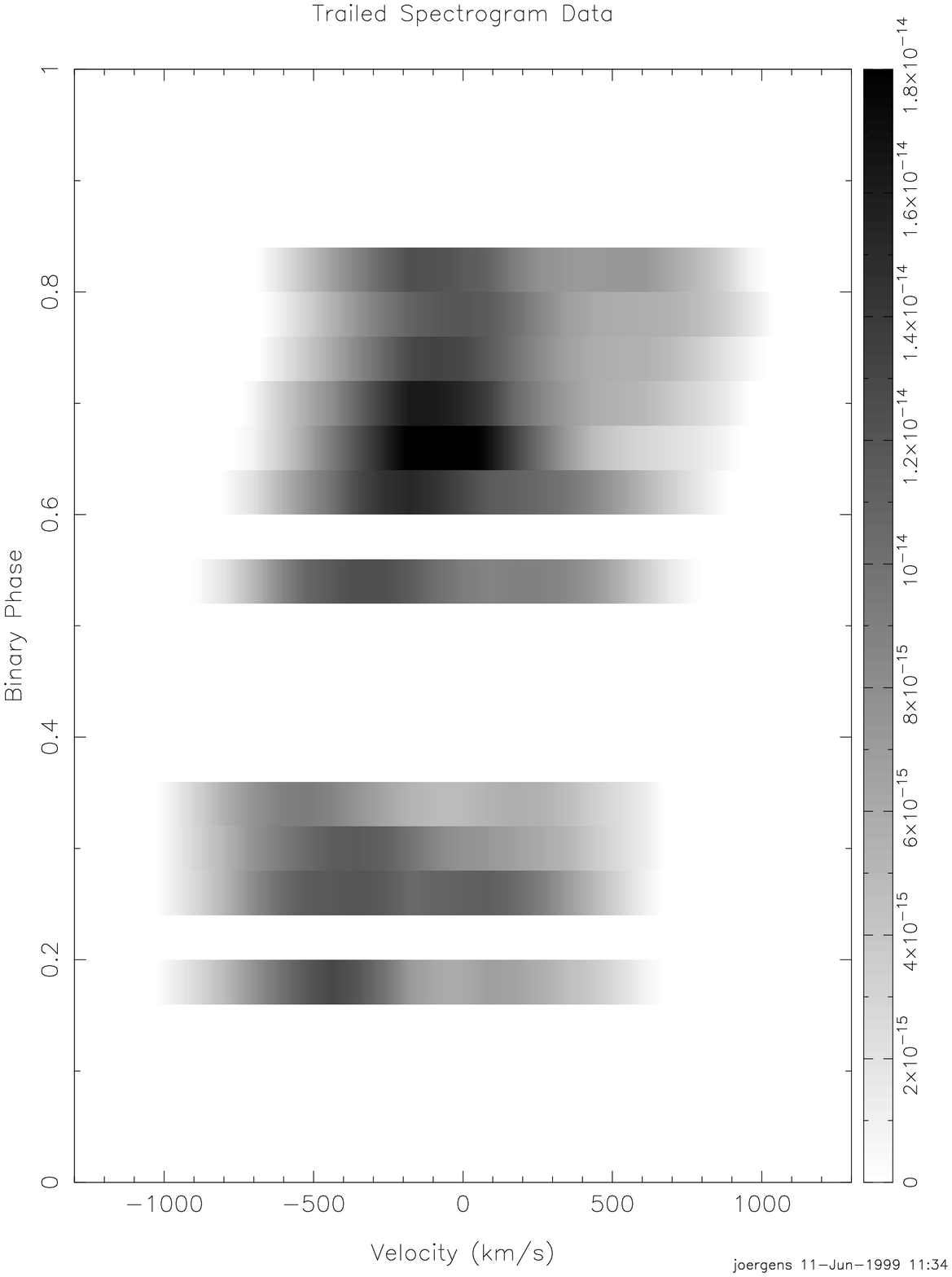}
\hfill\includegraphics[bbllx=14,bblly=26pt,bburx=525,bbury=720,clip,width=4cm]{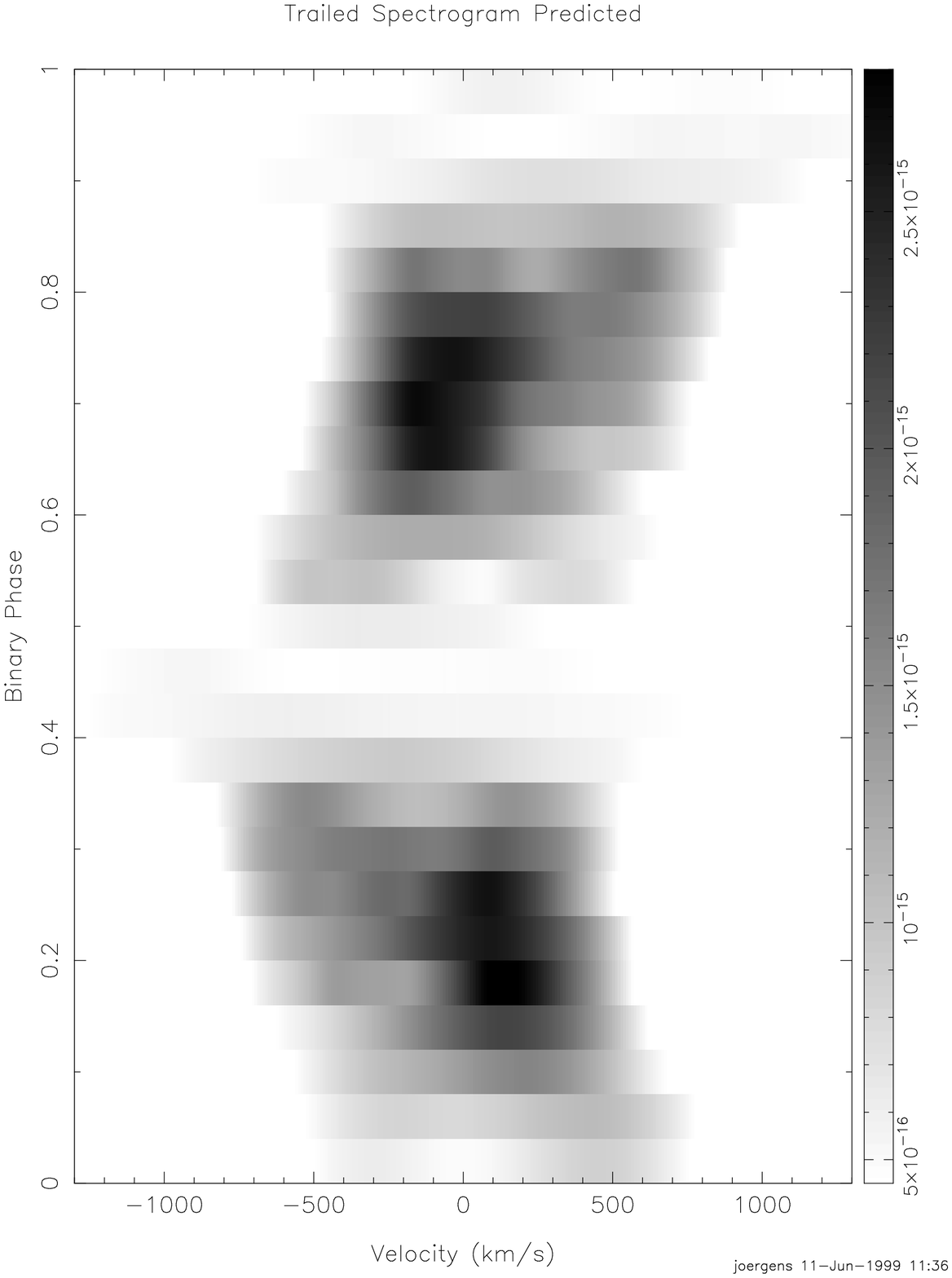}
\hfill\includegraphics[clip,width=5.2cm]{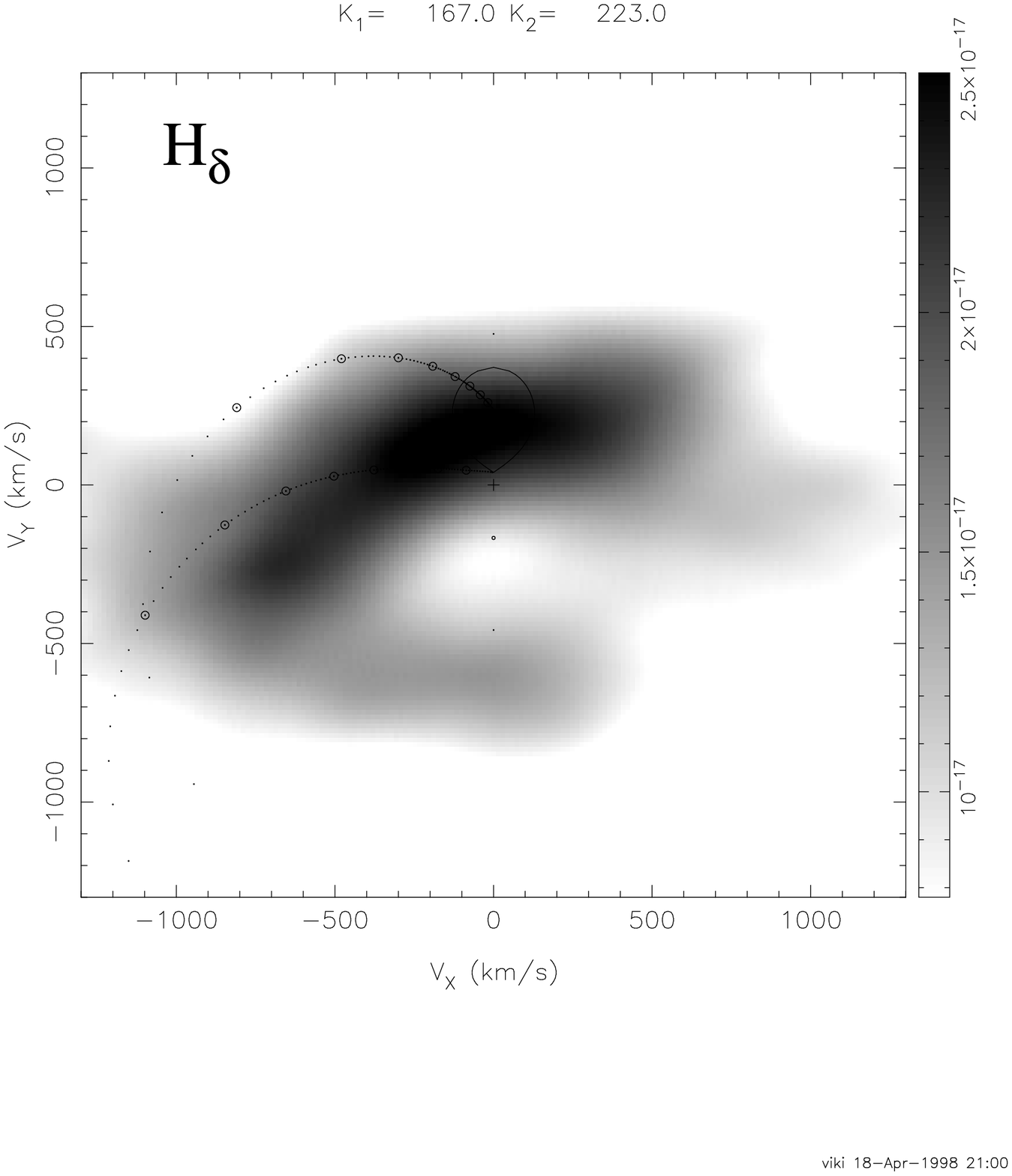}\hfill\mbox{}
}
\vspace{0.5cm} 
} 
\caption []{\label{outbmap1} \small Doppler maps of H$_{\alpha}$,
H$_{\beta}$ and H$_{\delta}$ during the outburst state of EX\,Dra
(right column), the observed phase--folded spectra 
(left column) and the spectra reconstructed from the maps (middle).
The images show
strong secondary star emission. Emission from the gas stream
obviously contributes to the bright emission region in the
H$_{\alpha}$ 
map. Due to an
incomplete phase coverage of the outburst data set an artificial
smearing in the V$_x$ direction occurs.
}
\end{figure*}

\begin{figure*}
\vbox{
\hbox{\mbox{}
\hfill\includegraphics[bbllx=14,bblly=26pt,bburx=525,bbury=720,clip,width=4cm]{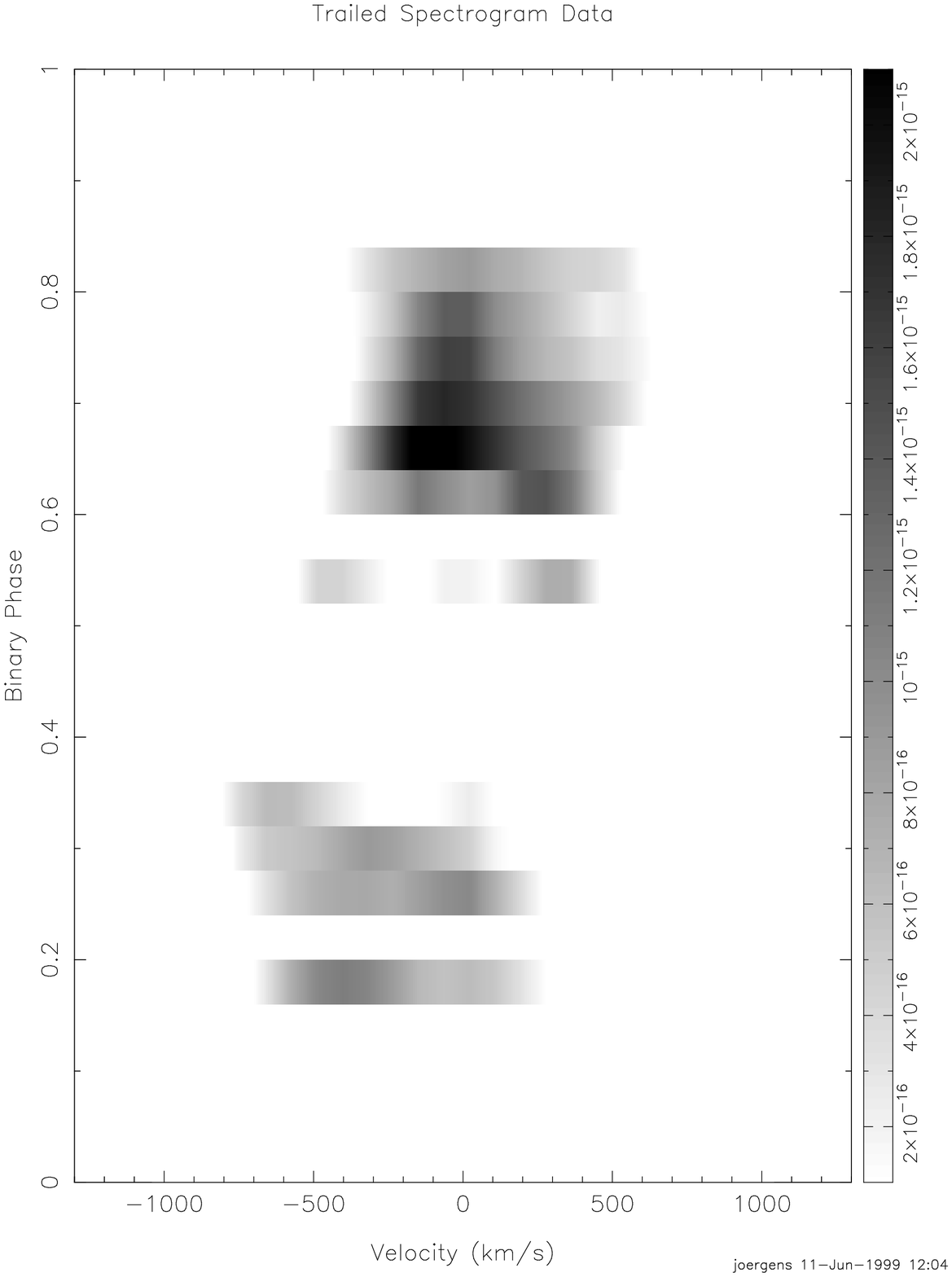}
\hfill\includegraphics[bbllx=14,bblly=26pt,bburx=525,bbury=720,clip,width=4cm]{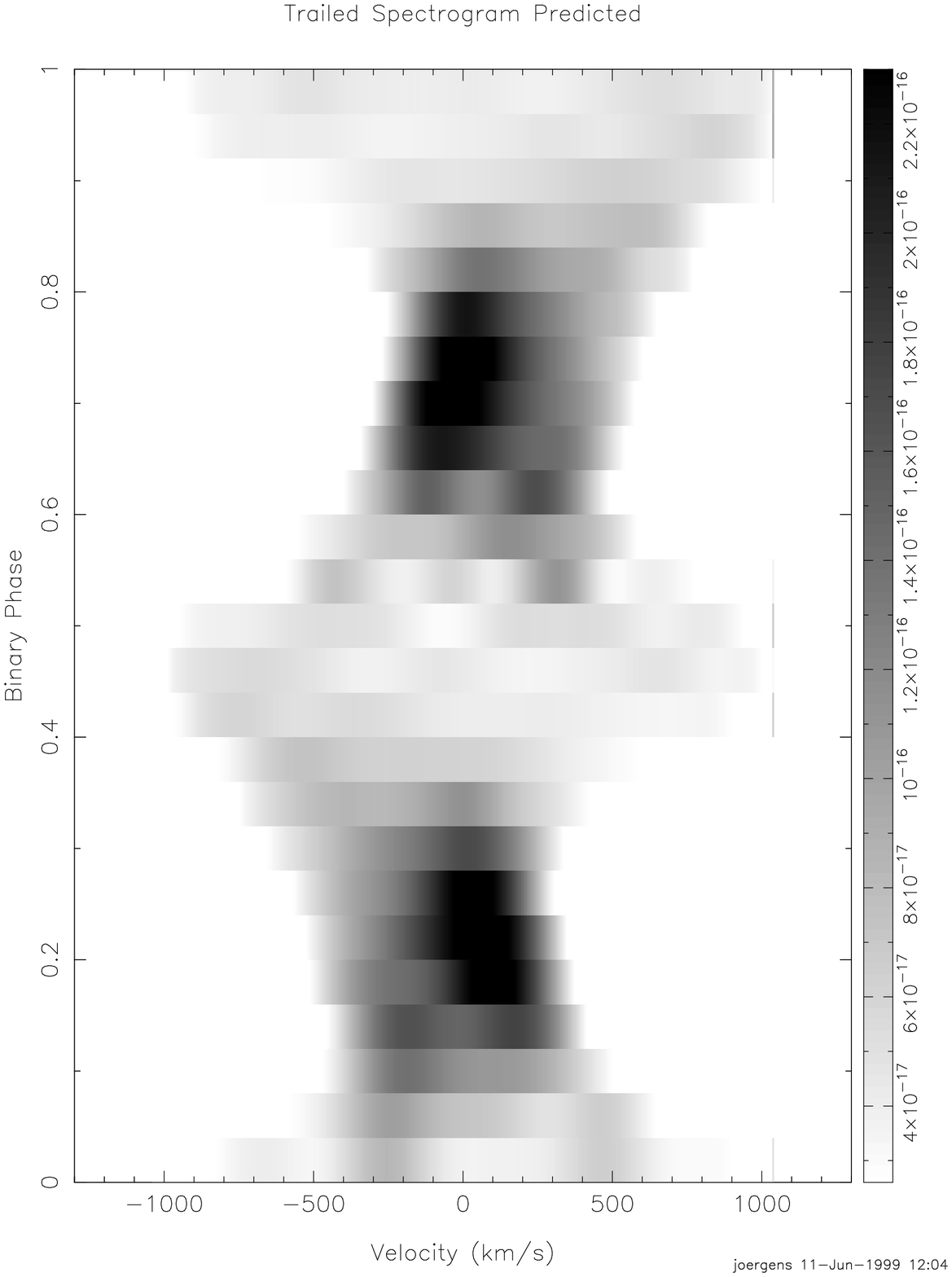}
\hfill\includegraphics[clip,width=5.2cm]{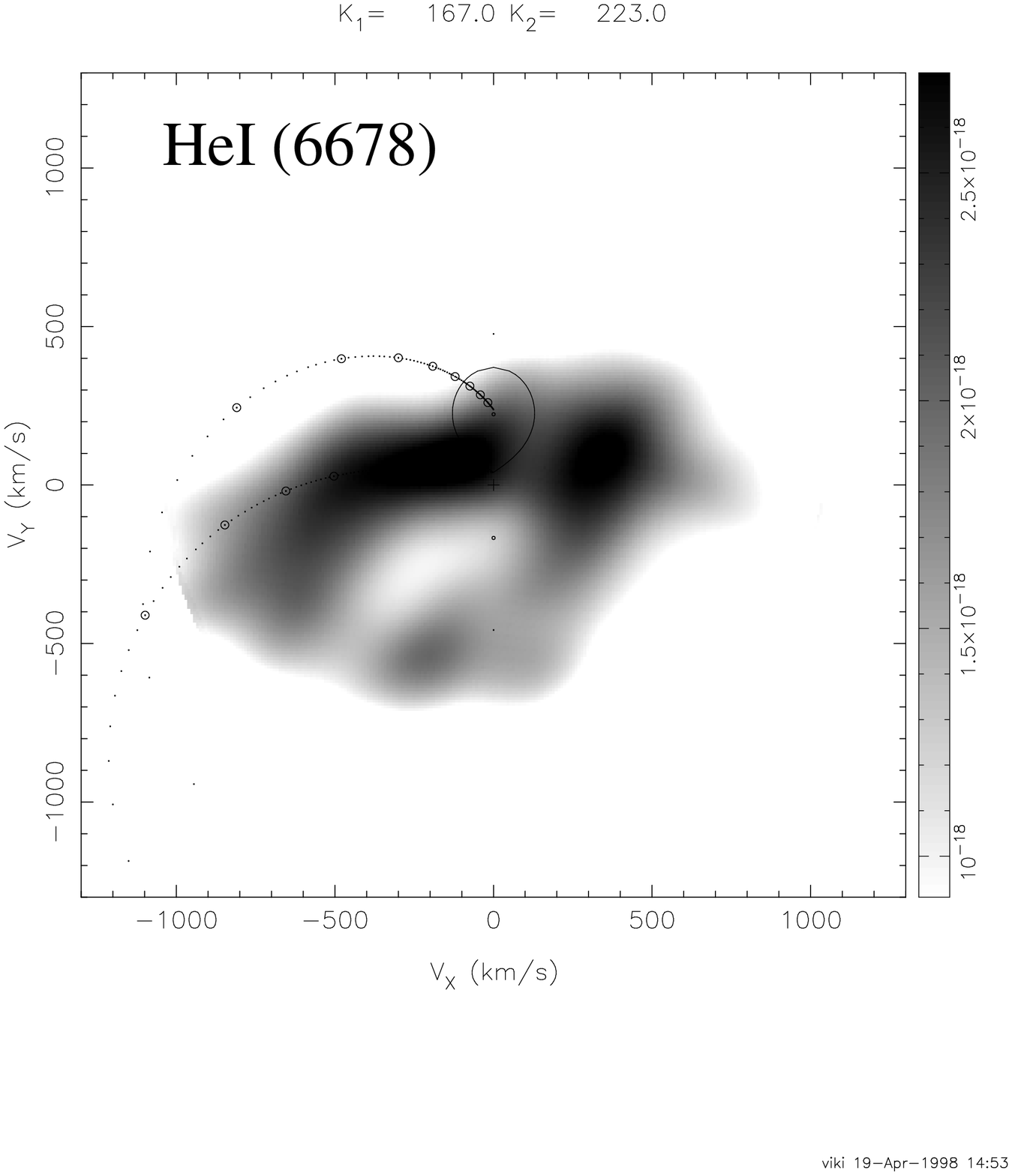}\hfill\mbox{}
}
\hbox{\mbox{}
\hfill\includegraphics[bbllx=14,bblly=26pt,bburx=525,bbury=720,clip,width=4cm]{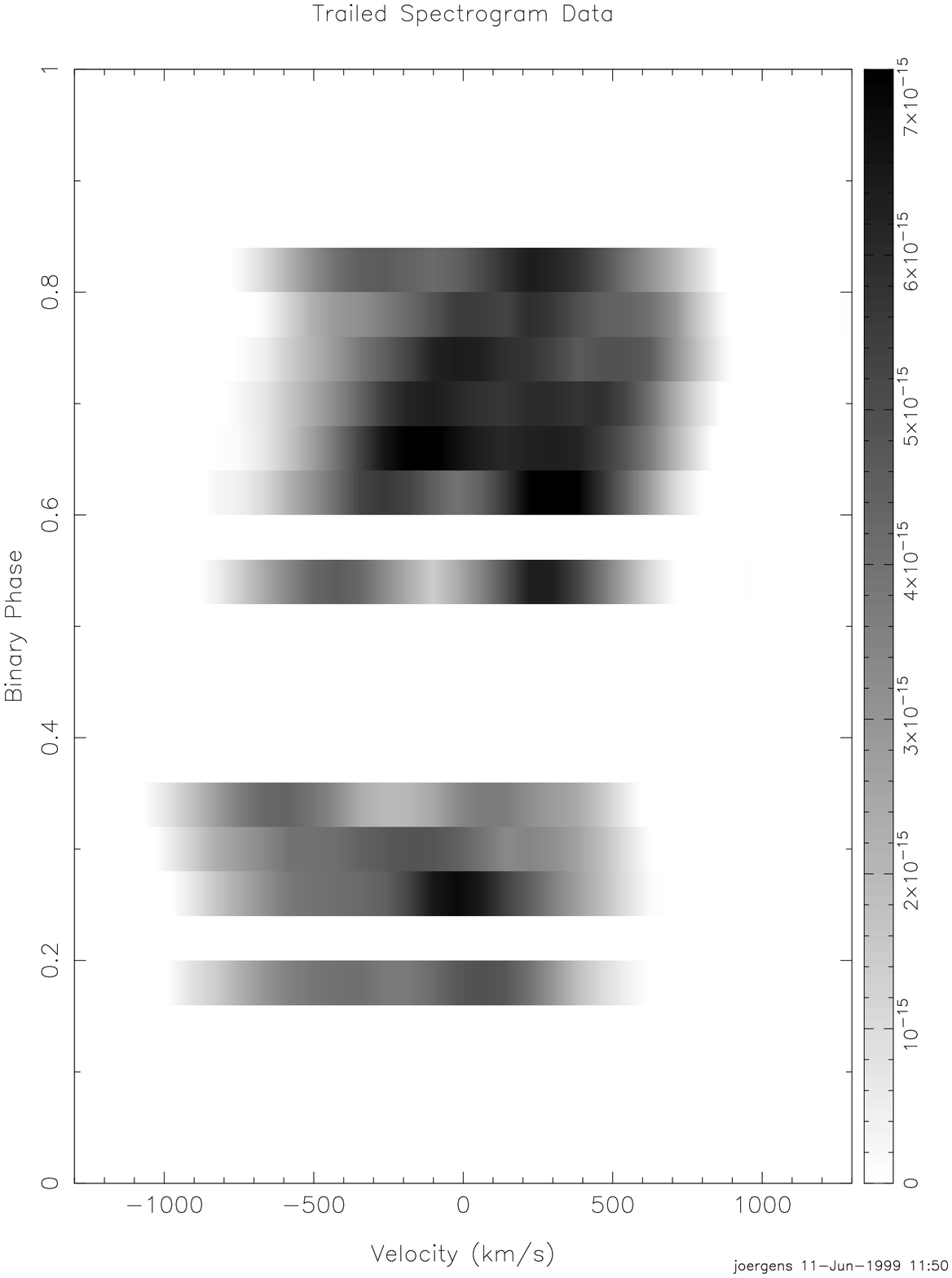}
\hfill\includegraphics[bbllx=14,bblly=26pt,bburx=525,bbury=720,clip,width=4cm]{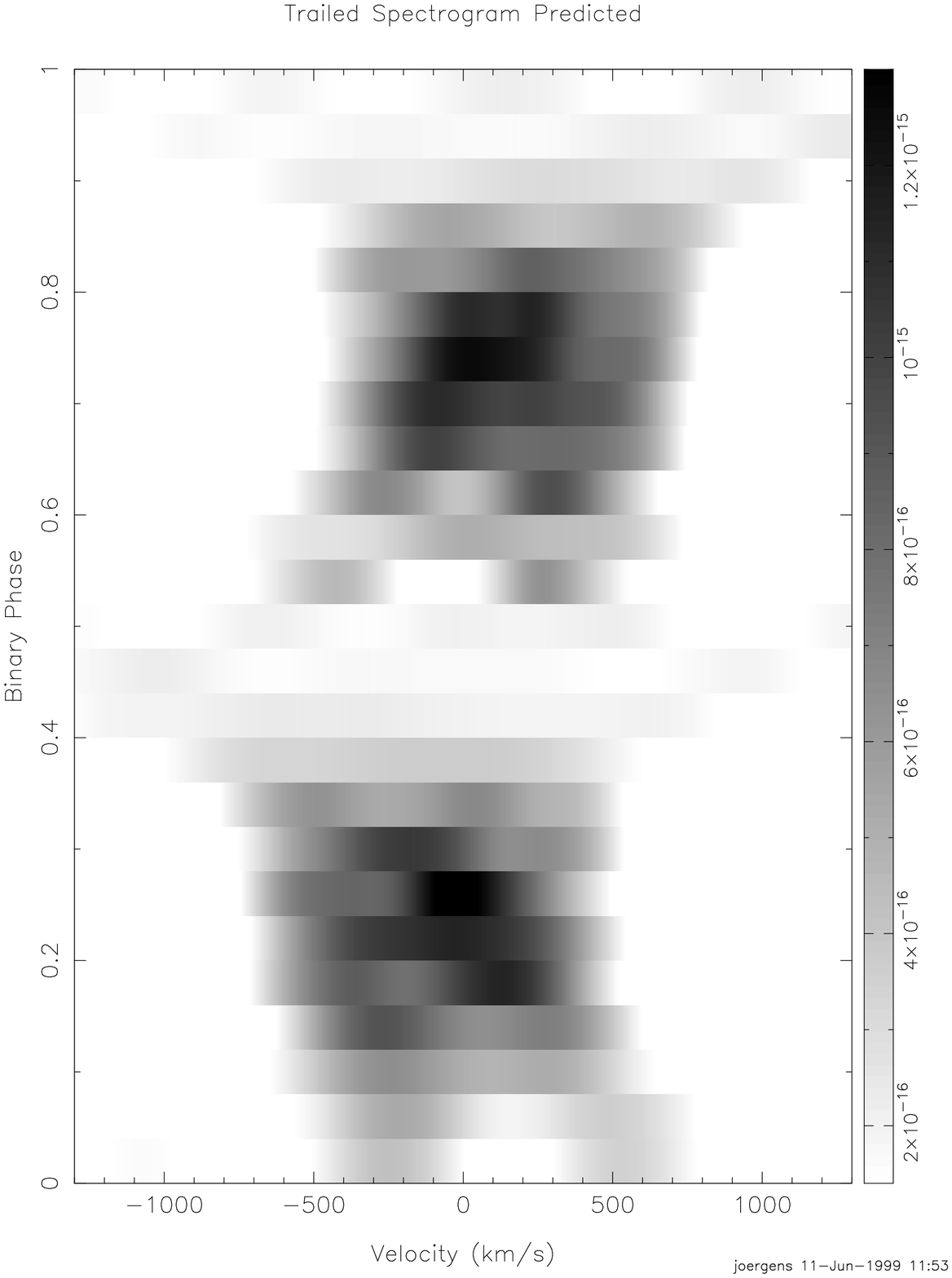}
\hfill\includegraphics[clip,width=5.2cm]{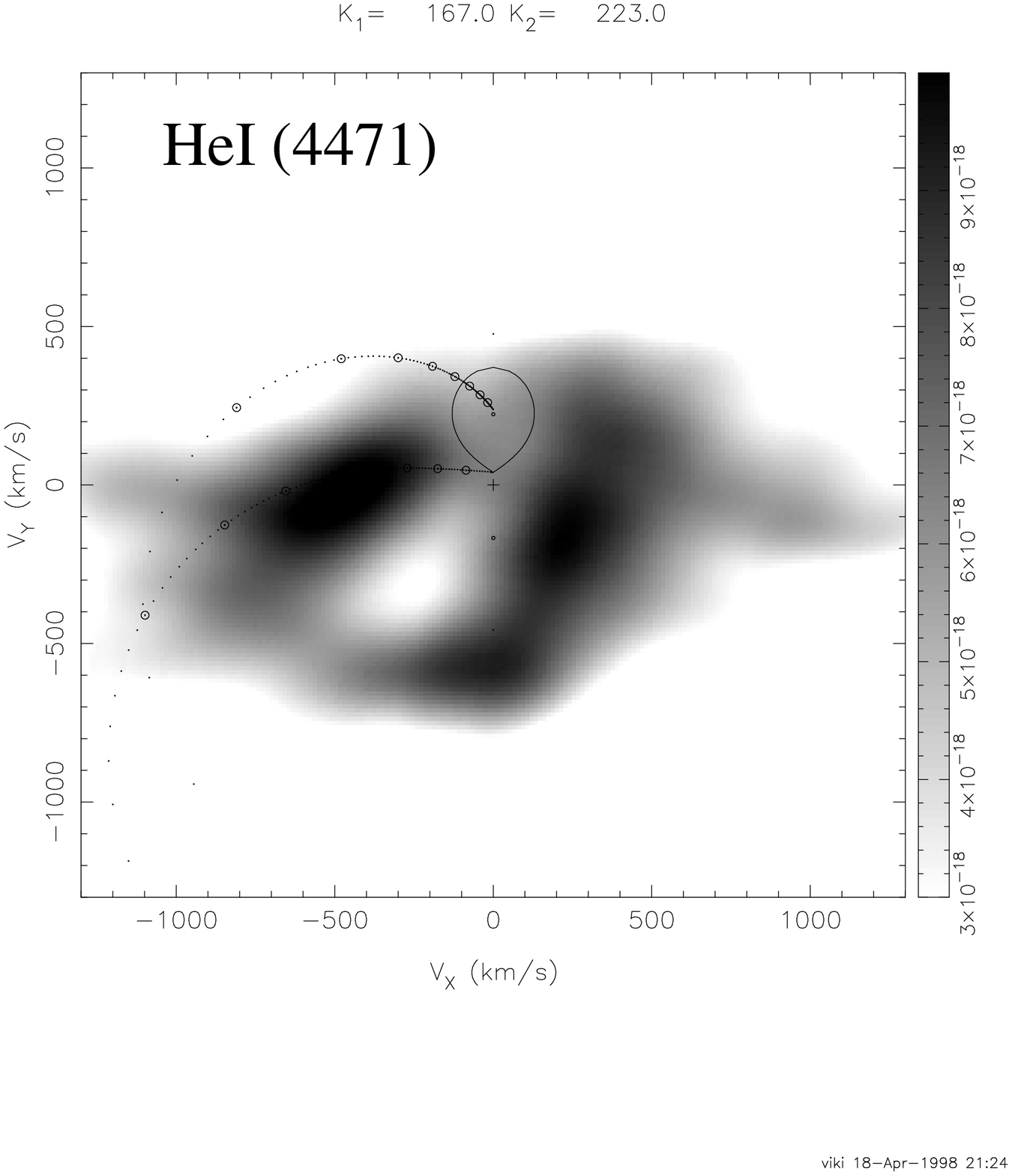}\hfill\mbox{}
}
%
\hbox{\mbox{}
\hfill\includegraphics[bbllx=14,bblly=26pt,bburx=525,bbury=720,clip,width=4cm]{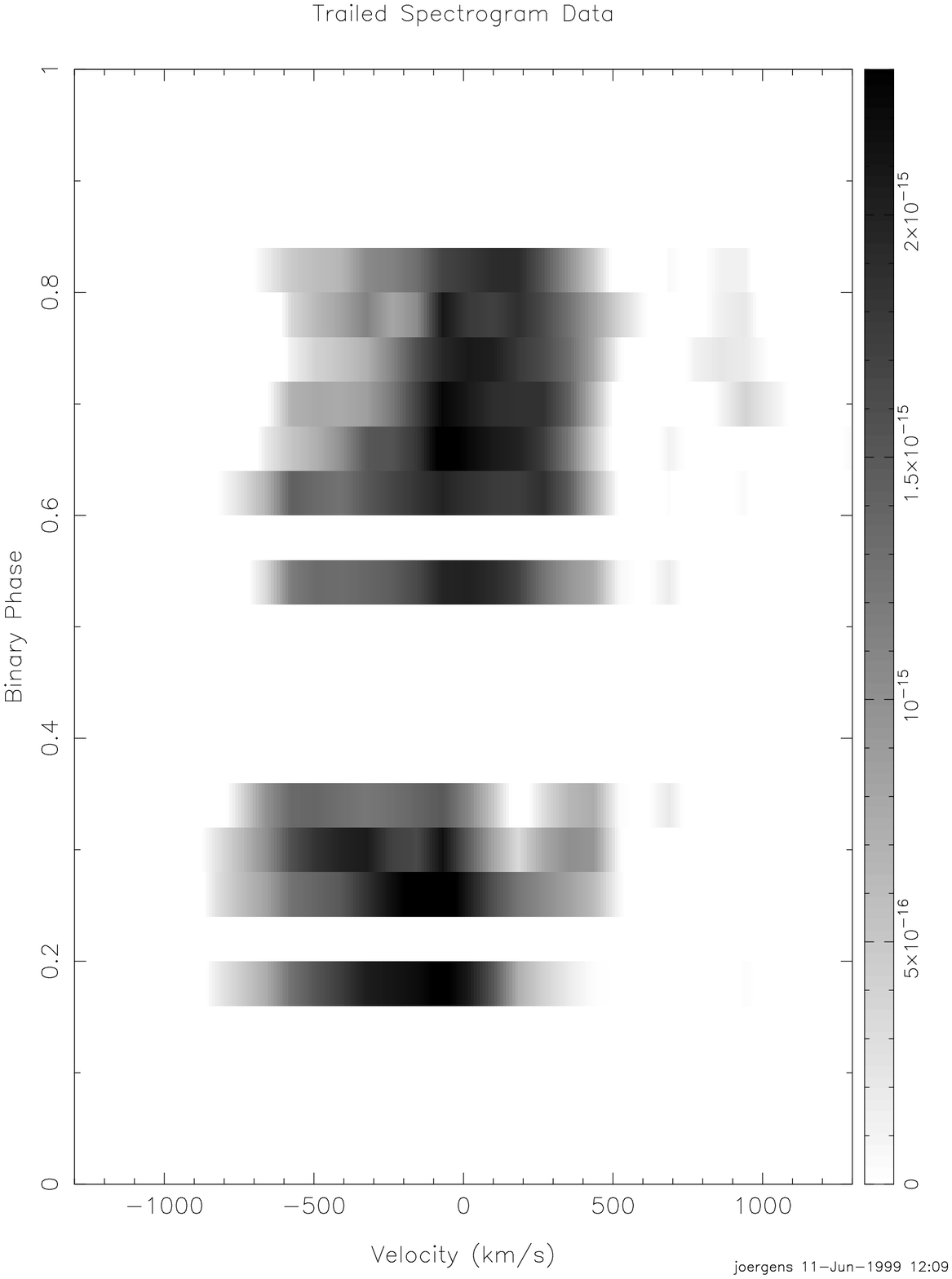}
\hfill\includegraphics[bbllx=14,bblly=26pt,bburx=525,bbury=720,clip,width=4cm]{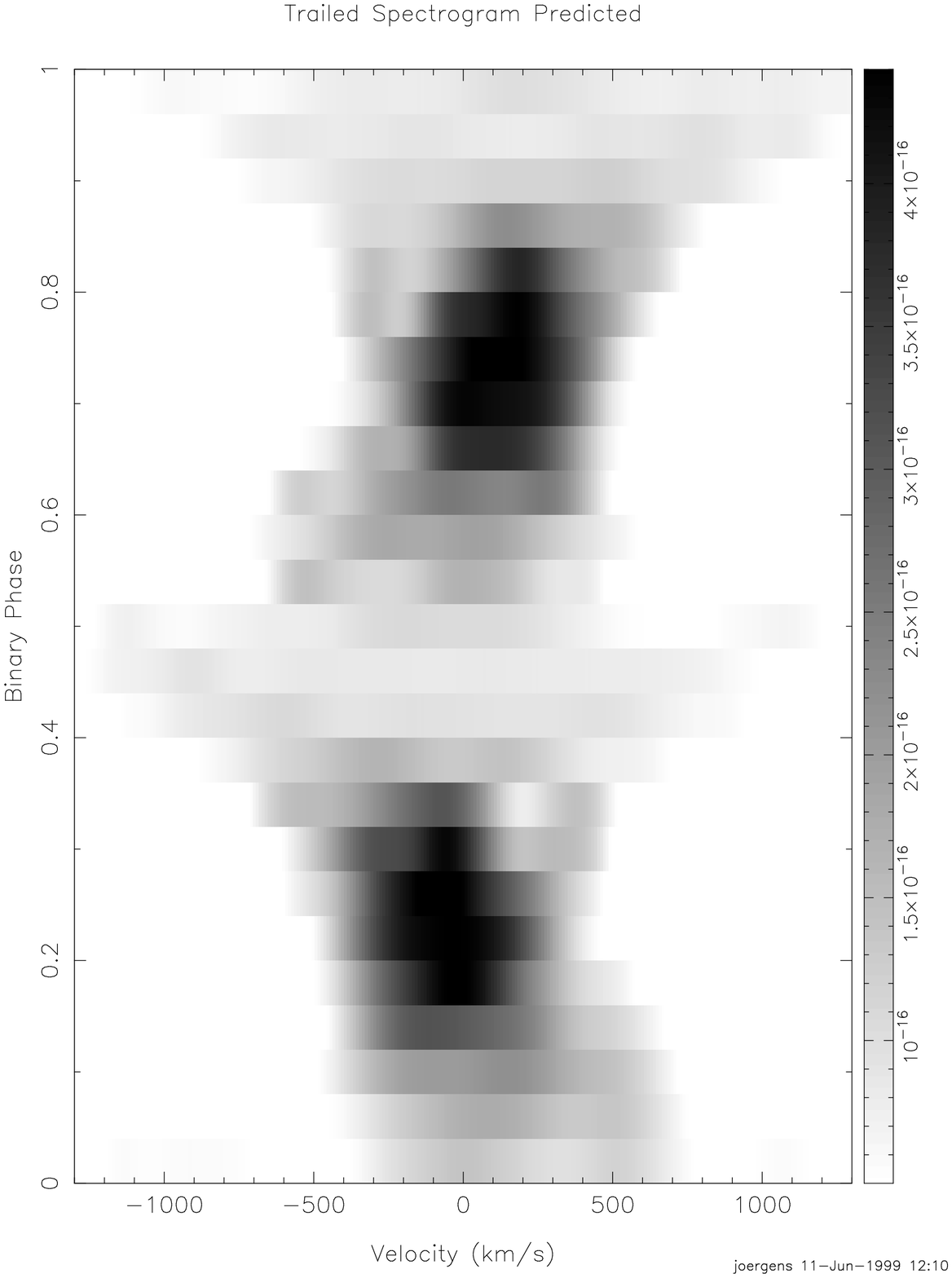}
\hfill\includegraphics[clip,width=5.2cm]{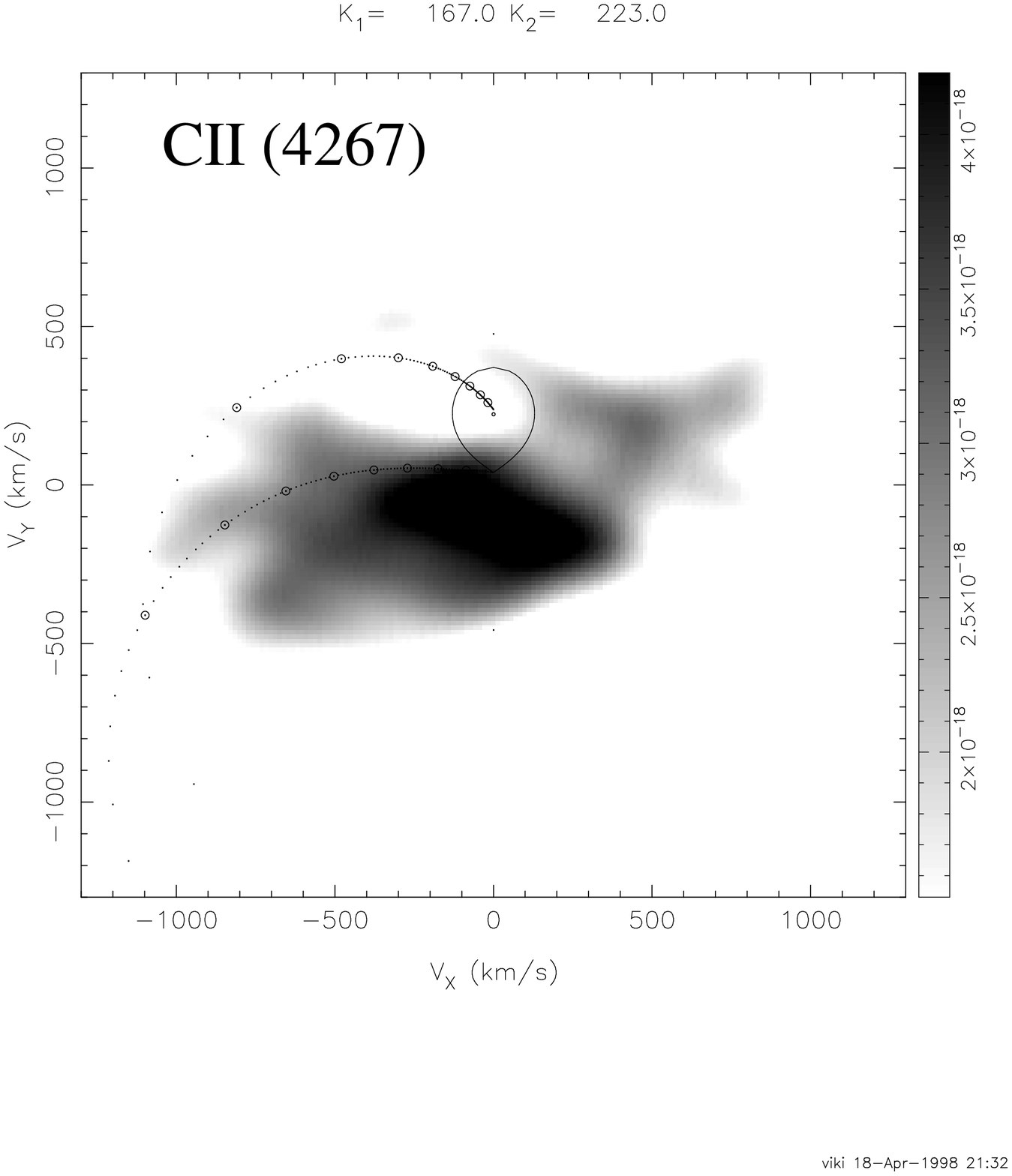}\hfill\mbox{}
}
}
\caption [] {\label{outbmap2} \small Doppler maps of 
\ion{He}{i} ($\lambda\lambda$ 6678, 4471\,{\AA}) and 
\ion{C}{ii} ($\lambda$ 4267\,{\AA}) during the outburst state of
EX\,Dra (right column), the observed phase--folded spectra 
(left column) and the spectra reconstructed from the maps (middle). 
In the \ion{He}{i} maps there is detectable emission from the gas 
stream and a further structure almost opposite to this feature.
The \ion{C}{ii} line differs totally from the other lines. A narrow component in the phase--folded data maps into a region in the velocity map close to the primary component.
}
\end{figure*}

We computed Doppler images for the emission lines H$_{\alpha}$,
H$_{\beta}$, H$_{\gamma}$, H$_{\delta}$, \ion{He}{i}
($\lambda\lambda$\,6678, 4471\,{\AA})
of the quiescence spectra and 
H$_{\alpha}$,
H$_{\beta}$, H$_{\delta}$, \ion{He}{i}
($\lambda\lambda$\, 6678, 4471\,{\AA}) and \ion{C}{ii}
($\lambda$ 4267\,{\AA}) of the outburst spectra. 
Fig.~\ref{quiescmap1} and Fig.~\ref{quiescmap2} show the results for the 
quiescent state and Fig.~\ref{outbmap1} and Fig.~\ref{outbmap2} those for the outburst state. For each line the phase--folded spectra (left column), the corresponding Doppler map (right column) and the spectra reconstructed from the map (middle) are displayed. 

The maps of H$_{\beta}$, H$_{\gamma}$ (Fig.~\ref{quiescmap1})
and H$_{\delta}$ (Fig.~\ref{quiescmap2})
during quiescence are dominated by a 
broad ring--like structure 
reflecting emission from the accretion disk centered around the white
dwarf. The projection of this ring along any direction produces a
double--peaked profile 
(middle column in Fig.~\ref{quiescmap1}, Fig.~\ref{quiescmap2})
corresponding to the observed double--peaked profiles 
(left column in Fig.~\ref{quiescmap1}, Fig.~\ref{quiescmap2})
seen in the Balmer lines at any given phase, with the 
exception of H$_{\alpha}$.

The outer edge of the disk seen in the H$_{\beta}$,
H$_{\gamma}$ and H$_{\delta}$ light of EX\,Dra is represented by the
inner edge of the 
ring--like structure in velocity coordinates, located
somewhere between 350 and 
500\,km\,s$^{-1}$. 

Emission from the gas stream
is detectable in all of the
Balmer lines as well as in the \ion{He}{i} ($\lambda\lambda$\,6678,
4471\,{\AA}) lines.
Except in H$_{\alpha}$ it is the brightest region in the emission
distributions. The series of small circles along the gas stream mark
the distance from the white dwarf at intervals of 0.1 R$_{L1}$
starting from R$_{L1}$. There is detectable emission between
0.5\,$\dots$0.3\, R$_{L1}$ in H$_{\beta}$,
H$_{\gamma}$, H$_{\delta}$ and \ion{He}{i} ($\lambda$\,4471\,{\AA}).

The quiescence data are quite noisy and the time resolution is not sufficient to resolve the obviously existing complex structures in the line profiles. However, an enhancement in intensity around phase 0.2 and 0.8 can be attributed to the bright spot emission. The corresponding S--wave is not visible during the whole binary orbit as it is for an anisotropically radiating emission site.
Doppler tomography is based on the assumption that the integrated flux is constant
with phase which is violated by any anisotropically emitting spot.
This leads to a discrepancy between observed and reconstructed
spectra.

Billington~et~al. (1996) also observed H$_{\alpha}$ emission from
the gas stream in EX\,Dra. The center of this emission is located at
V$_x\approx-500$\,km\,s$^{-1}$, which is in agreement with the center
of the gas stream emission in our quiescence H$_{\alpha}$ map. But we
do not observe the strong broadening of the gas stream emission
ranging over velocities of 1000\,km\,s$^{-1}$, detected by
Billington~et~al. (1996).

The H$_{\alpha}$ map deviates significantly
from that of the other Balmer lines and the \ion{He}{i} lines.
A second dominant emission spot is present in H$_{\alpha}$ at velocity
coordinates V$_x \approx 0$\dots$+500$\,km\,s$^{-1}$ and V$_y \approx
-200\dots-500$\,km\,s$^{-1}$. This feature is even brighter than the
one associated with the gas stream.
It is remarkable that a similar feature at about the same
velocity coordinates (although fainter) has also been 
detected in the H$_{\alpha}$ map of EX\,Dra performed by 
Billington~et~al. (1996).
Obviously this feature must be attributed to an emission site within
the disk opposite to the bright spot.

Furthermore,
Doppler tomography maps a small amount of
the H$_{\alpha}$ emission into the Roche lobe of the secondary star
(cp. Fig.~\ref{quiescmap1}). Imaging in velocity coordinates does
not allow to unequivocally distinguish whether this emission
originates at the secondary star or within the disk. This is a general
problem of Doppler tomography that two particles at different spatial
positions but
with the same Doppler velocities map into the same pixel in 
velocity space. Since the emission displays velocities smaller than
200\,km\,s$^{-1}$ where no disk emission is expected and since it is
visible on that side of the 
secondary facing the primary it could be caused by photospheric
heating of the secondary due to irradiation by the primary
component. The emission seems to be concentrated near the poles of the
irradiated side of the secondary and the L$_1$ point does not seem to
be affected.
This suggests the 
white dwarf or the boundary layer as the ionizing source  and a
shadowing of the equatorial parts, including the L$_1$ point by the
disk.

Billington~et~al. (1996) also found evidence for irradiation of the
secondary in their H$_{\alpha}$ Doppler map. In contrast to our result
they observed strong broadened secondary star emission
uniformly distributed over the secondary including L$_1$. 
This might be due to their incomplete phase
coverage and fewer number of spectra compared to our data set.

%


The disk structures in the
H$_{\gamma}$, H$_{\delta}$ and \ion{He}{i} ($\lambda$\,4471\,{\AA}) maps  
are superposed by several emission features weaker than the bright spot.
The features in H$_{\gamma}$ and H$_{\delta}$ are consistent. 
Since such features can be caused by a low phase resolution
they might be artificial. 
Further
spectroscopic observations with higher time and velocity resolution
are required
to verify their existence.


\subsection{Doppler imaging in the outburst state}

Interpretation of Doppler tomograms of the outburst state is hampered
by the fact that the outburst data set covers
only 56\% of the binary orbit, which could result in artificial
effects. Due to a non uniform distribution
of the available spectra across the binary orbit 
we obtain a decreased resolution in the V$_x$ direction in the maps.

As in many dwarf novae during outburst, EX\,Dra shows strong
Balmer emission from the secondary star in the high accretion
state (Fig.~\ref{outbmap1}). The secondary star displays no emission
of the \ion{He}{i} triplet 
line at $\lambda$\,4471\,{\AA} 
(Fig.~\ref{outbmap2})  
whereas there are indications of \ion{He}{i}
($\lambda$\,6678\,{\AA}) emission from the secondary during
outburst.

Because of the mentioned artificial smearing in the V$_x$ direction the
bright emission region in the H$_{\alpha}$ outburst map can be 
a superposition of emission from the irradiated
secondary and the gas stream.
Emission from the gas stream is further
detectable in the \ion{He}{i} maps, most prominently in the line 
at $\lambda$\,4471\,{\AA}. A comparison with the \ion{He}{i}
($\lambda$\,4471\,{\AA}) quiescence map 
locates the impact region in outburst about 0.2\,R$_{L1}$
closer to the L$_1$ point than in quiescence.
This can be attributed to the enhanced accretion
during outburst, which causes an enlargement of the outer
disk regions due to conservation of angular momentum within the disk.

The system seen in the light of the single ionized carbon
(Fig.~\ref{outbmap2}) differs totally from the other images. 
The \ion{C}{ii} emission which is only marginally detectable 
in quiescence is powerful enhanced during outburst. The S--wave in the 
observed phase--folded spectra (lower left picture in Fig.~\ref{outbmap2}) 
indicates that the 
\ion{C}{ii} line flux is emitted at relatively low radial velocities.
In the Doppler image the \ion{C}{ii} emission is 
concentrated in the center of the
disk suggesting a line forming region close to the white dwarf. Whether
\ion{C}{ii} originates in the chromosphere of the primary or in parts
of the inner boundary layer or in a wind from the white dwarf is not clear.
The reconstructed data (lower middle picture in Fig.~\ref{outbmap2}) indicate
a reliable back--projection, but the 
interpretation of the \ion{C}{ii} line is hampered by the low S/N
ratio of the data recorded within this line and by possible artificial
effects due to the incomplete phase coverage and the low phase resolution.
However, it is obvious that during outburst 
the \ion{C}{ii} line is not emitted from the same sites, where the
Balmer or helium lines originate.

The two \ion{He}{i} outburst maps (cp. Fig.~\ref{outbmap2}) clearly display 
emission features in the disk almost opposite to the bright spot.
Recently detected spiral structures in the outburst accretion disk of IP\,Peg 
(\cite{Steeghs97}) suggest that tidally induced spiral shocks 
(first proposed by \cite{Sawada86}, 1987) 
may 
also play a role in the accretion processes during outburst in other dwarf 
novae.
Unfortunately the low spectral and phase--resolution of our outburst
data permit no reliable statement about spiral structures in the disk of
EX\,Dra.
The detection of spiral patterns by means of 
Doppler tomography demands a spectral resolution of
$\sim$~80\,km\,s$^{-1}$ and a time resolution of $\sim$~40 spectra per
binary orbit (\cite{Steeghs99}), which is not met by our outburst
data set. 

\section[]{Discussion and conclusion}
\label{Conclusions}

EX\,Dra in quiescence is dominated by emission from a fully
established accretion disk and by emission from the gas stream interacting
with its outer rim. The center of the gas stream
emission located at 
$ 
\mbox{V}_x \approx -500\,\mbox{km}\,\mbox{s}^{-1} 
$ 
during quiescence
is in agreement with that in the H$_{\alpha}$ map of
Billington~et~al. (1996), but our data set does not confirm the strong
broadening detected by Billington~et~al (1996). 

Unlike most other dwarf novae, the emission lines of EX\,Dra
remain strong during outburst. This behaviour is also seen in
other deeply eclipsing dwarf novae like
IP\,Peg (Marsh \& Horn 1990), Z\,Cha (\cite{Vogt82}) and OY\,Car
(\cite{laDous91}). The high orbital inclination of these systems
probably accounts for 
the strong emission lines during eruption, because at large
inclinations the flux in the optically thick continuum of the accretion disk is
reduced by projection and limb darkening in favor of
emission lines formed above the disk.
%

Reemission of the secondary star in EX\,Dra is detectable
during quiescence in H$_{\alpha}$ and during outburst in H$_{\alpha}$,
H$_{\beta}$ and H$_{\delta}$, where it becomes the dominating emission
source during outburst. The emission is concentrated near the poles of
that side of the secondary facing the primary and indicates photospheric
heating caused by irradiation by the white dwarf or the boundary layer.

Emission lines from highly excited species, like \ion{He}{ii}
($\lambda$\,4686\,{\AA}), \ion{C}{ii} ($\lambda$\,4267\,{\AA}) and
\ion{C}{iii}/\ion{N}{iii} 
($\lambda\lambda$\,4634 $\dots$ 4651\,{\AA}) are
only marginally or not at all
detected during quiescence but become strongly enhanced during outburst.

The Doppler image of the \ion{C}{ii} line during outburst
locates its line forming region close to the white dwarf.
This suggests the chromosphere of the white dwarf or the inner
boundary layer or an outflow as possible emission sites. 
Radial velocity measurements
of this line should in principle reproduce the radial velocity of the
primary with higher
precision than the distorted disk emission lines do. However, due to
the low
S/N of the \ion{C}{ii} line and the incomplete phase coverage of the
recorded outburst spectra, we were not able to determine the radial
velocity of the \ion{C}{ii} line.

The \ion{He}{ii} line is superimposed by \ion{C}{iii}/\ion{N}{iii}
emission, for that reason no reliable Doppler map of this line can be
performed. Provided that the mass transfer rate
$\dot{\mbox{M}}(\mbox{d})$ exceeds 10$^{-9}\,\mbox{M}_{\sun}/\mbox{y}$
(\cite{Patterson85}) the \ion{He}{ii} emission during outburst might be 
produced by reprocessing of soft X--rays from the boundary layer in
the disk.

It is further conceivable that \ion{He}{ii} (and \ion{C}{ii}) is formed
in outflowing material, but this would require a very slow wind. However,
the material emitting the \ion{He}{ii} and \ion{C}{ii} lines can not
be too extended, but must be closely confined to the orbital plane
since both emission lines are eclipsed at phase $\phi = 0$.

A comparison between the location of the gas stream emission in
the Doppler maps 
of \ion{He}{i} ($\lambda$\,4471\,{\AA}) in quiescence and outburst 
show that the disk radius is enlarged by about 0.2\,R$_{L1}$
during outburst.

The brightest line forming region in quiescence in the Doppler map
of H$_{\alpha}$ is located far from the gas stream trajectory. We
have no explanation which line forming processes are involved to
produce this feature.
Since this emission spot is only detectable in the H$_{\alpha}$ line
and not in lines reflecting higher temperatures it is unlikely to be
caused by a second impact region,
as discussed e.g. by 
Lubow (1989).
This emission spot is partly responsible for the
complex H$_{\alpha}$ line profiles and the
asymmetric profile in the averaged spectra in form of an enhanced
red--shifted peak.

The \ion{He}{i} maps computed for the outburst spectra show evi\-dence
for emission structures in the right and top right region of the
Doppler maps. Unfortunately the quality of the outburst
  spectra does not allow to draw a clear decision
  whether these structures may be attributed to
  spiral structures within the accretion disk of
  EX Dra, as detected for IP\,Peg during
outburst (\cite{Steeghs97}), or not.

Our results show that EX\,Dra is a very interesting CV and that the
next step should be spectroscopy with higher time and spectral
resolution. The fact that the system can be frequently found in the
outburst state should make further investigation possible throughout it's
outburst cycle.

\begin{acknowledgements}
      We thank 
      Vadim Burwitz for helpful discussions and useful comments on the
      manuscript. We also thank 
      C. Ries for collecting photometric data at Wendelstein observatory.
      Part of this work was supported by the
      \emph{Deut\-sche For\-schungs\-ge\-mein\-schaft}, grant
      Ba\,867/3-1, Ba\,867/5-1. 
\end{acknowledgements}



\begin{thebibliography}{}

   \bibitem[Barwig et al. 1987]{Barwig87} Barwig H., Schoembs R.,
      Buckenmayer C., 1987, A\&A 175, 327 

   \bibitem[Barwig et al. 1993]{Barwig93} Barwig H., Fiedler H.,
      Reimers D., Bade N., 1993, 
      in: Compact Stars in Binary Systems,
      ed.\ H. van Woerden, Abstracts of IAU Symp.\ 165, p.\ 89

   \bibitem[Billington et al. 1996]{Billington96} Billington I., Marsh
      T.R., Dhillon V.S., 1996, MNRAS 278, 673 

   \bibitem[Fiedler et al. 1997]{Fiedler97} Fiedler H., Barwig H.,
        Mantel K.H., 1997 A\&A, 327, 173 
  
   \bibitem[Horne 1986]{Horne86} Horne K., 1986,
	PASP 98, 609


   \bibitem[Horne 1991]{Horne91} Horne K., 1991, In: Fundamental
        Properties of Cataclysmic Variable Stars: 12th North American
        Workshop on Cataclysmic Variables and Low Mass X--ray
        Binaries, San Diego State University Publication, San Diego,
        ed. A.W. Shafter, p. 160

   \bibitem[la Dous 1991]{laDous91} la Dous C., 1991, A\&A 252, 100 

   \bibitem[Lubow 1989]{Lubow89} Lubow S.H., 1989, ApJ 340, 1064 

   \bibitem[Marsh \& Horne 1988]{Marsh88} Marsh T.R., Horne K., 1988,
        MNRAS 235, 269 

   \bibitem[Marsh \& Horne 1990]{Marsh90} Marsh T.R., Horne K., 1990, 
        ApJ 349, 593 

   \bibitem[Patterson \& Raymond 1985]{Patterson85} Patterson J.,
        Raymond J.C., 1985, ApJ 292, 550 

   \bibitem[Sawada et al. 1986]{Sawada86} Sawada K., Matsuda T., Hachisu
   I., 1986, MNRAS 219, 75

   \bibitem[Sawada et al. 1987]{Sawada87} Sawada K., Matsuda T., Inoue
   M., Hachisu I., 1987, MNRAS 224, 307

   \bibitem[Steeghs \& Stehle 1999]{Steeghs99} Steeghs D., Stehle R.
       1999 MNRAS 307, 99

   \bibitem[Steeghs et al. 1997]{Steeghs97} Steeghs D., Harlaftis
       E.T., Horne K., 1997 MNRAS 270, L28, Erratum: 1998 MNRAS 296, 463 

   \bibitem[Vogt 1982]{Vogt82} Vogt N., 1982, ApJ 252, 563

   \bibitem[Wolf et al. 1998]{Wolf98} Wolf S., Barwig H., Bobinger A.,
        Mantel K.H., Simic D., 1998, A\&A 332, 984

\end{thebibliography}
\end{document}